\documentclass[aps, prd, reprint,showpacs,  preprintnumbers,amsmath,amssymb,superscriptaddress, nofootinbib]{revtex4-1}
\usepackage{amssymb, amsmath, braket, nicefrac, graphicx} 
\usepackage{relsize}
\usepackage{lineno}
\usepackage{setspace}

\usepackage{ esint }
\usepackage{chngpage}
\usepackage{epsfig}
\usepackage{multirow}
\usepackage{rotating}
\usepackage{float}
\usepackage{color}
\usepackage{xcolor}
\usepackage{appendix}
\usepackage{ amstext }
\usepackage{ gensymb }
\usepackage{ textcomp }
\usepackage{ wasysym }
\usepackage{pstricks}

\usepackage[caption=false,labelformat=parens]{subfig}

\usepackage{dcolumn}
\usepackage{bm}
\usepackage{units}

\newcommand{\numu}{\mbox{$\nu_{\mu}$}}                   

\newcommand{\anu}{\ensuremath{\bar{\nu}}}

\newcommand{\anumu}{\ensuremath{\bar{\nu}_{\mu}}}

\newcommand{\pip}{\mbox{$\pi^{+}$}}       
\newcommand{\piz}{\mbox{$\pi^{0}$}} 
\newcommand{\simgt}{\,\hbox{\lower0.6ex\hbox{$\sim$}\llap{\raise0.6ex\hbox{$>$}}}\,}
\newcommand{\simlt}{\,\hbox{\lower0.6ex\hbox{$\sim$}\llap{\raise0.6ex\hbox{$<$}}}\,}

\newcommand{\minerva}{MINERvA~}
\newcommand{\minos}{MINOS~}

\newcommand{\numubar}{\ensuremath{\bar{\nu}_{\mu}}~}

\definecolor{maroon}{RGB}{162,10,10}

\usepackage{hyperref}
\usepackage[all]{hypcap}
\hypersetup{
    colorlinks,
    citecolor=black,
    filecolor=black,
    linkcolor=black,
    urlcolor=black
}

\makeatletter

\renewenvironment{figure}
  {\def\@captype{figure}}
  {}
\makeatother

\begin{document}
\preprint{FERMILAB-PUB-16-228-ND}
\title{Cross sections for $\numu$ and $\anumu$ induced pion production on hydrocarbon\\ in the few-GeV region using \minerva}

\newcommand{\Rutgers}{Rutgers, The State University of New Jersey, Piscataway, New Jersey 08854, USA}
\newcommand{\Hampton}{Hampton University, Dept. of Physics, Hampton, VA 23668, USA}
\newcommand{\Dortmund}{Institute of Physics, Dortmund University, 44221, Germany }
\newcommand{\Otterbein}{Department of Physics, Otterbein University, 1 South Grove Street, Westerville, OH, 43081 USA}
\newcommand{\JMU}{James Madison University, Harrisonburg, Virginia 22807, USA}
\newcommand{\Florida}{University of Florida, Department of Physics, Gainesville, FL 32611}
\newcommand{\UCIrvine}{Department of Physics and Astronomy, University of California, Irvine, Irvine, California 92697-4575, USA}
\newcommand{\CBPF}{Centro Brasileiro de Pesquisas F\'{i}sicas, Rua Dr. Xavier Sigaud 150, Urca, Rio de Janeiro, Rio de Janeiro, 22290-180, Brazil}
\newcommand{\PUCP}{Secci\'{o}n F\'{i}sica, Departamento de Ciencias, Pontificia Universidad Cat\'{o}lica del Per\'{u}, Apartado 1761, Lima, Per\'{u}}
\newcommand{\INRM}{Institute for Nuclear Research of the Russian Academy of Sciences, 117312 Moscow, Russia}
\newcommand{\Jlab}{Jefferson Lab, 12000 Jefferson Avenue, Newport News, VA 23606, USA}
\newcommand{\Pittsburgh}{Department of Physics and Astronomy, University of Pittsburgh, Pittsburgh, Pennsylvania 15260, USA}
\newcommand{\Guanajuato}{Campus Le\'{o}n y Campus Guanajuato, Universidad de Guanajuato, Lascurain de Retana No. 5, Colonia Centro, Guanajuato 36000, Guanajuato M\'{e}xico.}
\newcommand{\Athens}{Department of Physics, University of Athens, GR-15771 Athens, Greece}
\newcommand{\Tufts}{Physics Department, Tufts University, Medford, Massachusetts 02155, USA}
\newcommand{\WM}{Department of Physics, College of William \& Mary, Williamsburg, Virginia 23187, USA}
\newcommand{\FNAL}{Fermi National Accelerator Laboratory, Batavia, Illinois 60510, USA}
\newcommand{\Purdue}{Department of Chemistry and Physics, Purdue University Calumet, Hammond, Indiana 46323, USA}
\newcommand{\MCLA}{Massachusetts College of Liberal Arts, 375 Church Street, North Adams, MA 01247}
\newcommand{\UMD}{Department of Physics, University of Minnesota -- Duluth, Duluth, Minnesota 55812, USA}
\newcommand{\Northwestern}{Northwestern University, Evanston, Illinois 60208}
\newcommand{\UNI}{Universidad Nacional de Ingenier\'{i}a, Apartado 31139, Lima, Per\'{u}}
\newcommand{\Rochester}{University of Rochester, Rochester, New York 14627 USA}
\newcommand{\Austin}{Department of Physics, University of Texas, 1 University Station, Austin, Texas 78712, USA}
\newcommand{\USM}{Departamento de F\'{i}sica, Universidad T\'{e}cnica Federico Santa Mar\'{i}a, Avenida Espa\~{n}a 1680 Casilla 110-V, Valpara\'{i}so, Chile}
\newcommand{\Geneva}{University of Geneva, 1211 Geneva 4, Switzerland}
\newcommand{\Chicago}{Enrico Fermi Institute, University of Chicago, Chicago, IL 60637 USA}
\newcommand{\hired}{}
\newcommand{\OregonState}{Department of Physics, Oregon State University, Corvallis, Oregon 97331, USA}
\newcommand{\oxford}{}
\newcommand{\bmeThanks}{now at SLAC National Accelerator Laboratory, Stanford, CA 94309, USA}
\newcommand{\higueraThanks}{now at University of Houston, Houston, TX 77204, USA}
\newcommand{\damartinezThanks}{now at Illinois Institute of Technology, Chicago, IL 60616, USA}
\newcommand{\mcgivernThanks}{now at Fermi National Accelerator Laboratory, Batavia, IL 60510, USA}
\newcommand{\joelmousseauThanks}{now at University of Michigan, Ann Arbor, MI 48109, USA}
\newcommand{\LazaThanks}{also at Department of Physics, University of Antananarivo, Madagascar}
\newcommand{\twaltonThanks}{now at Fermi National Accelerator Laboratory, Batavia, IL 60510, USA}
\newcommand{\jwolcottThanks}{now at Tufts University, Medford, MA 02155, USA}
\newcommand{\zavalaThanks}{Deceased}

\author{C.L.~McGivern}\thanks{\mcgivernThanks}  \affiliation{\Pittsburgh}
\author{T.~Le}                            \affiliation{\Tufts}  \affiliation{\Rutgers}
\author{B.~Eberly}\thanks{\bmeThanks}     \affiliation{\Pittsburgh}
\author{L.~Aliaga}                        \affiliation{\WM}  \affiliation{\PUCP}
\author{O.~Altinok}                       \affiliation{\Tufts}
\author{L.~Bellantoni}                    \affiliation{\FNAL}
\author{A.~Bercellie}                     \affiliation{\Rochester}
\author{M.~Betancourt}                    \affiliation{\FNAL}
\author{A.~Bodek}                         \affiliation{\Rochester}
\author{A.~Bravar}                        \affiliation{\Geneva}
\author{H.~Budd}                          \affiliation{\Rochester}
\author{T.~Cai}                           \affiliation{\Rochester}
\author{M.F.~Carneiro}                    \affiliation{\CBPF}
\author{M.E.~Christy}                     \affiliation{\Hampton}
\author{H.~da~Motta}                      \affiliation{\CBPF}
\author{S.A.~Dytman}                      \affiliation{\Pittsburgh}
\author{G.A.~D\'{i}az~}                   \affiliation{\Rochester}  \affiliation{\PUCP}
\author{E.~Endress}                       \affiliation{\PUCP}
\author{J.~Felix}                         \affiliation{\Guanajuato}
\author{L.~Fields}                        \affiliation{\FNAL}  \affiliation{\Northwestern}
\author{R.~Fine}                          \affiliation{\Rochester}
\author{R.Galindo}                        \affiliation{\USM}
\author{H.~Gallagher}                     \affiliation{\Tufts}
\author{T.~Golan}                         \affiliation{\Rochester}  \affiliation{\FNAL}
\author{R.~Gran}                          \affiliation{\UMD}
\author{D.A.~Harris}                      \affiliation{\FNAL}
\author{A.~Higuera}\thanks{\higueraThanks}  \affiliation{\Rochester}  \affiliation{\Guanajuato}
\author{K.~Hurtado}                       \affiliation{\CBPF}  \affiliation{\UNI}
\author{M.~Kiveni}                        \affiliation{\FNAL}
\author{J.~Kleykamp}                      \affiliation{\Rochester}
\author{M.~Kordosky}                      \affiliation{\WM}
\author{E.~Maher}                         \affiliation{\MCLA}
\author{S.~Manly}                         \affiliation{\Rochester}
\author{W.A.~Mann}                        \affiliation{\Tufts}
\author{C.M.~Marshall}                    \affiliation{\Rochester}
\author{D.A.~Martinez~Caicedo}\thanks{\damartinezThanks}  \affiliation{\CBPF}
\author{K.S.~McFarland}                   \affiliation{\Rochester}  \affiliation{\FNAL}
\author{A.M.~McGowan}                     \affiliation{\Rochester}
\author{B.~Messerly}                      \affiliation{\Pittsburgh}
\author{J.~Miller}                        \affiliation{\USM}
\author{A.~Mislivec}                      \affiliation{\Rochester}
\author{J.G.~Morf\'{i}n}                  \affiliation{\FNAL}
\author{J.~Mousseau}\thanks{\joelmousseauThanks}  \affiliation{\Florida}
\author{D.~Naples}                        \affiliation{\Pittsburgh}
\author{J.K.~Nelson}                      \affiliation{\WM}
\author{A.~Norrick}                       \affiliation{\WM}
\author{Nuruzzaman}                       \affiliation{\Rutgers}  \affiliation{\USM}
\author{V.~Paolone}                       \affiliation{\Pittsburgh}
\author{J.~Park}                          \affiliation{\Rochester}
\author{C.E.~Patrick}                     \affiliation{\Northwestern}
\author{G.N.~Perdue}                      \affiliation{\FNAL}  \affiliation{\Rochester}
\author{L.~Rakotondravohitra}\thanks{\LazaThanks}  \affiliation{\FNAL}
\author{M.A.~Ramirez}                     \affiliation{\Guanajuato}
\author{R.D.~Ransome}                     \affiliation{\Rutgers}
\author{H.~Ray}                           \affiliation{\Florida}
\author{L.~Ren}                           \affiliation{\Pittsburgh}
\author{D.~Rimal}                         \affiliation{\Florida}
\author{P.A.~Rodrigues}                   \affiliation{\Rochester}
\author{D.~Ruterbories}                   \affiliation{\Rochester}
\author{H.~Schellman}                     \affiliation{\OregonState}  \affiliation{\Northwestern}
\author{D.W.~Schmitz}                     \affiliation{\Chicago}  \affiliation{\FNAL}
\author{C.~Simon}                         \affiliation{\UCIrvine}
\author{C.J.~Solano~Salinas}              \affiliation{\UNI}
\author{S.~S\'{a}nchez~Falero}            \affiliation{\PUCP}
\author{B.G.~Tice}                        \affiliation{\Rutgers}
\author{E.~Valencia}                      \affiliation{\Guanajuato}
\author{T.~Walton}\thanks{\twaltonThanks}  \affiliation{\Hampton}
\author{J.~Wolcott}\thanks{\jwolcottThanks}  \affiliation{\Rochester}
\author{M.Wospakrik}                      \affiliation{\Florida}
\author{D.~Zhang}                         \affiliation{\WM}

\date{\today}

\begin{abstract}
Separate samples of charged-current pion production events representing two
semi-inclusive channels $\numu$-CC($\pi^{+}$) and $\anumu$-CC($\pi^{0}$) have
been obtained using neutrino and antineutrino exposures of the \minerva detector.   
Distributions in kinematic variables based upon $\mu^{\pm}$-track 
reconstructions are analyzed and compared for the two samples.   The
differential cross sections for muon production angle, muon momentum, 
and four-momentum transfer $Q^2$,  are reported, 
and cross sections versus neutrino energy are obtained.   Comparisons 
with predictions of current neutrino event generators are used to clarify
the role of the $\Delta(1232)$ and higher-mass baryon resonances in CC pion 
production and to show the importance of pion final-state interactions. 
For the $\numu$-CC($\pi^{+}$) ($\anumu$-CC($\pi^{0}$)) sample, the absolute data rate
is observed to lie below (above) the predictions of some of the event generators by amounts
that are typically 1-to-2\,$\sigma$.   However the generators are able 
to reproduce the shapes of the differential cross sections 
for all kinematic variables of either data set.
\end{abstract}

\maketitle


\section{Introduction}
Interactions of few-GeV neutrinos and antineutrinos with nuclei are of keen interest to present and future
neutrino oscillation experiments, such as T2K, NOvA, DUNE, 
and HyperKamiokande~\cite{T2K-expt, NOvA-expt, DUNE-expt, HyperK-expt}.
In this energy region,  charged-current single-pion production (CC($\pi$))  competes 
with quasielastic scattering in terms of the total charged-current (CC) event rate observed in the 
near and far detectors of the neutrino oscillation experiments.
In $\numu/\anumu$ CC scattering on nuclei, the nuclear medium 
enables directly-produced CC($\pi$) states to morph into other
final-state pion channels and into quasielastic-like scattering topologies 
as well.   These cross-channel migrations involve energy transfer
from the produced state to the struck nucleus, rendering the total final-state energy difficult to detect.  
In this way distortions are introduced into the reconstruction of
neutrino energy $E_{\nu}$,  four-momentum transfer squared $Q^2$,  and hadronic invariant mass $W$.
Obtaining precise knowledge of the observed CC($\pi$) rates and relating them to the various ways that 
directly-produced states can feed into the final states actually observed,  
is crucial for continued progress in neutrino oscillation measurements.

Two previous publications~\cite{Brandon-pion, Trung-pion} reported
the MINERvA experiment's first measurements 
of CC pion production on hydrocarbon (CH) in the channels
\begin{equation}
\label{reaction-1}
  \numu  + \text{CH}  \rightarrow \mu^{-} + n\pi^{\pm} + X, 
\end{equation}
\begin{equation}
\label{reaction-2}
  \anumu  + \text{CH}  \rightarrow  \mu^{+} + \pi^{0} + X'.
\end{equation}

\noindent
For both of these CC reactions it is possible to reconstruct the incident neutrino energy, $E_\nu$, 
the squared four-momentum transfer to the struck nucleus, $Q^2$, 
and the invariant hadronic mass, $W$.  Through event selection, 
the charged pion sample of process \eqref{reaction-1} is dominated by \pip~production. 
The data in Ref.~\cite{Brandon-pion} was presented in two 
different ways - a single-pion sample with $W<$ 1.4\,GeV and
an n-pion sample with $W<$ 1.8\,GeV where n signifies one or more
charged pions.  The sample selection for the latter sample (same as the data presented here)
is for a semi-inclusive process; $X$ may include, in addition to the recoil nucleon, 
neutral pions, and other particles (nucleons and photons) released by nuclear de-excitation and final-state interactions (FSI).   
The neutral pion sample of reaction~\eqref{reaction-2} is more nearly exclusive~\cite{Trung-pion}.
The sample is restricted to events having one and only one $\pi^0$, with no visible charged tracks other than the 
$\mu^{+}$ emerging from the primary vertex.  The recoil system $X'$ is limited to the recoil nucleon plus
de-excitation neutrons and photons.  There is no limitation on the value of $W$.

For the analysis reported here, the selected event samples for reactions \eqref{reaction-1} and \eqref{reaction-2}
are restricted by requiring all events to have hadronic mass $W$ less than 1.8 GeV and neutrino
energy in the range 1.5\,GeV $<E_{\nu}<$ 10\,GeV.   Here, $W$ is calculated from the true muon kinematics and true $E_{\nu}$.
Consequently the charged pion sample is nearly identical
to the n-pion sample in Ref.~\cite{Brandon-pion} and the neutral pion sample is slightly smaller than in Ref.~\cite{Trung-pion}.
The restriction on final-state hadronic mass serves to enhance the contribution of $\Delta(1232)$ and $N^{*}$ resonance production
relative to that from CC DIS processes.    Moreover the hadronic mass selection, together with the requirement
that a Michel electron be observed on a non-muon track from the primary vertex, isolates 
a subsample of process \eqref{reaction-1} that is more nearly a $\pi^{+}$ production sample, as will be elaborated below. 

The two separate CC pion production event samples were obtained with
the NuMI beam in the low-energy mode, with the 
horn-current focusing set to produce a beam of predominantly $\numu$ or $\bar{\nu}_\mu$. Consequently, 
the spectral shapes and effective $E_{\nu}$ range of the initiating $\numu/\anumu$ fluxes 
are similar for the two data sets.    The initial studies measured the rates and 
kinematic distributions for the produced pions.   Comparisons were made with generator predictions, 
and trends involving final-state interactions of the pions within the target carbon nuclei were identified.
These measurements have also been compared to
a phenomenological treatment of neutrino-induced pion production carried out
within the GiBUU transport theoretical framework~\cite{mosel-minerva-pion}.

In the present work,  the two CC($\pi$) event samples are investigated further and in tandem, enabling 
the scope of Refs.~\cite{Brandon-pion} and \cite{Trung-pion} to be significantly extended.  The present analysis encompasses 
the differential distributions of the final-state muon and of kinematic variables that are determined by
the muon kinematics, with $E_{\nu}$ and $Q^2$ receiving particular attention.   
The resulting measurements are complementary to the pion kinematical distributions previously
presented~\cite{Brandon-pion,Trung-pion}.   While the distributions of these previous works show
interesting sensitivity to the FSI processes, the distributions presented here depend on the combination of 
underlying pion-production reactions on single nucleons with nuclear medium effects
arising from nucleon-momentum distribution and nucleon-nucleon correlations.

Comparisons of muon-related kinematic distributions are used to elicit similarities and differences between the
$\nu$-induced and $\anu$-induced pion production datasets.   To illuminate the contributing processes,
each data distribution is also compared to predictions obtained using neutrino event generators.   
For the latter data-vs-simulation comparisons,
the analysis makes use of three neutrino event generators that are widely used by neutrino experiments, 
namely GENIE 2.6.2~\cite{Andreopoulos201087}, NEUT 5.3.3~\cite{ref:NEUT}, and NuWro~\cite{ref:NuWro}.   
These codes have been independently constructed and validated;  a summary of the phenomenological 
strategies and models used by each generator is given in Ref.~\cite{Brandon-pion}.

The measurements of this work utilize event selections and improved flux estimations that differ from those used by
Refs.~\cite{Brandon-pion, Trung-pion}.   These modifications are discussed in Secs. II and III.

\section{Overview of the Experiment}
 \subsection{Beam, Detector, and Exposures}
\label{subsec:B-D-E}

MINERvA uses a fine-grained, plastic-scintillator tracking 
detector~\cite{minerva_nim} in conjunction with the magnetized 
MINOS near detector~\cite{Michael:2008bc}, to record interactions 
of neutrinos and antineutrinos from the high-intensity NuMI beam 
at Fermilab~\cite{Anderson:1998-Adamson:2015}. 

The measurements reported here use the MINERvA detector central tracking 
volume as the target, with the surrounding electromagnetic and hadronic 
calorimeters providing containment.   The central volume has a hexagonal 
cross section of 2\,m inner diameter, extends longitudinally for 
2.5\,m, and has a mass of 5400\,kg.  It consists of planes 
of polystyrene scintillator strips oriented perpendicular to the 
horizontal axis of the detector.  The horizontal axis is inclined 
3.3$^\circ$  relative to the beam axis.  There are three scintillator-plane 
orientations, at 0$^\circ$ and $\pm 60^\circ$  relative to the detector 
vertical axis, that provide  X, U, and V ``views'' of interactions in 
the scintillator medium.  The module planes alternate between UX and 
VX pairs,  enabling 3-D reconstruction of vertices, charged tracks, and 
electromagnetic showers of neutrino events.   Separation of multiple 
interactions within a single 10\,$\mu$s beam spill is made possible by the 3.0\,ns 
timing resolution of the readout electronics.  

The MINOS near detector, located 2\,m downstream, serves as the muon 
spectrometer for the MINERvA central tracker.  A muon exiting downstream 
of MINERvA is tracked by the magnetized, steel-plus-scintillator planes 
of MINOS, enabling its momentum and charge to be measured.  The 
combination of position, angle, and timing in each detector allows 
matching of muon tracks in the two detectors.   Full descriptions of 
the design, calibration, and performance of the MINERvA detector 
configuration are available in Refs.~\cite{minerva_nim, Aliaga:2015aqe}. 

The data were taken between October 2009 and April 2012 using the 
low-energy NuMI mode, which produces a wide-band beam with neutrino energies 
extending from 1\,GeV to greater than 20 GeV and a peak energy of 
3\,GeV.   The current-polarity of the magnetic horns in the 
beamline is set to focus either $\pi^+$ or $\pi^-$, providing $\nu_\mu$ or $\bar{\nu}_\mu$
fluxes of approximately 92\% purity or 40\% purity respectively.  The $\numu$ 
CC charged-pion production events were obtained from an integrated 
exposure of $3.04 \times 10^{20}$ protons on target (POT); the $\numubar$ 
CC single-$\piz$ production events were obtained in exposures with a total 
of $2.01 \times 10^{20}$ POT. Half of the $\bar{\nu}_\mu$ exposure was taken 
with only the downstream half of the detector during construction. The two datasets 
are analyzed separately and the final results combined.

The $\numu$ and $\anumu$ fluxes for 
these exposures were calculated using a detailed simulation
of the NuMI beamline based on Geant4~\cite{Agostinelli2003250,1610988} and
constrained by published proton-carbon yield 
measurements~\cite{Alt:2006fr, Barton:1982dg, Lebedev:2007zz}.  Compared 
to the previous studies, Refs.~\cite{Brandon-pion, Trung-pion}, the 
measurements reported here benefit from improved flux predictions 
resulting from new constraints  based upon $\nu + e^{-}$ elastic scattering 
data from \minerva and from incorporation of new data 
on pion and kaon yields~\cite{Park-thesis, minerva-flux-2015}.
Consequently the present work contains improved estimations for absolute 
event rates for each of the two data sets.  All event-generator predictions 
are based upon the improved flux predictions.   Updates to the previously published
results~\cite{Brandon-pion, Trung-pion} using the new fluxes are presented in the Appendix
of this paper.

\subsection{Neutrino interaction modeling}
Neutrino and  antineutrino-nucleus interactions are simulated 
using version 2.6.2 of the GENIE neutrino event generator~\cite{Andreopoulos201087}.    
The generation of inelastic CC neutrino-nucleus interactions involves three different considerations: 
\begin{itemize}
\item{ {\it Target nucleons:}   
Nucleons inside the nucleus are treated as a relativistic Fermi gas.  
The nucleon momentum distribution is augmented with a high-momentum tail~\cite{bodekritchie} 
in order to account for short-range correlations.  The possibility 
of neutrino interactions on correlated-nucleon pairs is not included. }

\item {{\it The primary interaction:} 
Neutrino-induced pion production arising from a single struck nucleon
can proceed either by baryon-resonance excitation or by non-resonant processes. 
The baryon-resonance pion production is simulated using the Rein-Sehgal model~\cite{Rein:1980wg} 
with modern baryon resonance properties~\cite{pdg} and with 
an axial-vector mass of $M_{A}^{Res}=1.12\pm0.22$~GeV~\cite{Kuzmin:2006dh}. 
For non-resonant pion production, GENIE uses the Bodek-Yang model~\cite {Bodek:2004pc} 
with parameters adjusted to reproduce the neutrino-deuterium bubble chamber measurements 
over the final-state invariant hadronic mass range $W < 1.7$~GeV.}
The GENIE implementation does not include the Rein-Sehgal treatment of baryon-resonance interference, 
nor does it carry along the lepton mass terms in the cross section calculations.  

\item{ {\it Intranuclear interactions of final-state hadrons:} 
Final-state interactions (FSI) of hadrons produced 
inside the nucleus with the nuclear medium are simulated.  
The FSI are especially important for pions because 
of the very large pion-nucleon cross sections 
in the $\Delta(1232)$ resonance region.   In GENIE, an effective model 
for the FSI simulation is used in lieu of a full intranuclear cascade treatment.   That is,
pions can have at most one interaction on their way out of the nucleus~\cite{Dytman:2011zz}. 
This approximation works well for light nuclei such as carbon, 
the dominant target nucleus reported in this work. 
It is assumed that the pions produced inside the nucleus 
have the same pion-nucleus cross sections as beam pions, and so
scattering data from beam-pion measurements~\cite{Lee:2002eq,Ashery:1981tq} 
are used to model the interaction. The total interaction probability is determined 
by the pion-carbon total cross section.   An interaction of a pion 
within the nucleus proceeds by one of four processes, namely
pion absorption, inelastic scattering, elastic scattering, and charge exchange 
with probabilities according to the corresponding data.  
Kinematic distributions in the final state are set by algorithms 
that are fit to the corresponding pion-nucleus data.  }
\end{itemize}

 Coherent pion production is different from other interactions because 
 the neutrino interacts with the whole nucleus at once.   
 Coherent single-pion production by CC interactions on carbon has been
 measured by \minerva~\cite{ref:CC-Coh-Minerva}.
In GENIE, it is simulated according to the Rein-Sehgal model, 
updated with lepton mass terms~\cite{Rein:2006di}.

\subsection{Detector response}

Simulation of the response of the \minerva detector to particle propagation 
is provided by a Geant4-based model~\cite{Agostinelli2003250,1610988}.  
The scale for muon dE/dx energy loss in the detector 
is known to within 2\%.   The scale is established 
by requiring agreement between data and simulation for the reconstructed
energy deposited by momentum-analyzed, through-going muons. 
Hadron interactions in the detector materials are handled
by the Geant4 QGSP\textunderscore BERT physics list.

In order to reconstruct the energy of hadronic showers imaged by the detector, calorimetric corrections are required.
The procedure whereby these corrections are determined from simulation is described in 
Ref.~\cite{minerva_nim}.  A scaled-down replica of the MINERvA detector, operated in a low-energy
particle test beam, was used to establish the spectrometer's tracking efficiency and 
energy response to single hadrons, and to set the value for the Birks' constant of the scintillator~\cite{Aliaga:2015aqe}.
The average deviation between data and GEANT4 for pions was 5\%.

\section{CC Event selections and reconstruction} 

The 10\,$\mu$s NuMI beam spill is divided into ``time slices" that, 
based on the total visible energy in the scintillator as a function of time, encompass single events.
As a charged particle traverses the scintillator strips of the detector, its trajectory is recorded as individual
energy deposits (hits) having a specific charge content and time-of-occurrence.    The hits are grouped in time, and
then neighboring hits in each scintillator plane are gathered into objects called clusters.
 Clusters having more than 1\,MeV of energy are then matched among the three views to create a
track.   The per-plane position resolution is 2.7\,mm and the track
angular resolution is better than 10\,mrad~\cite{minerva_nim} in each view.    

A track that exits via the downstream surface of the \minerva spectrometer and matches with
a negatively-charged (positively-charged) track entering the
front of MINOS near detector, is taken to be the $\mu^- (\mu^+$) track of a CC event.
The reconstruction of the muon tracks in this experiment 
(including both \minerva and MINOS detectors) gives
a typical momentum resolution of $6\%$.  Muon track reconstruction incurs a small inefficiency due to event pileup.
This effect is studied by isolating individual tracks in either of the \minerva or \minos detectors, projecting them to the 
other detector, and then measuring the rate of reconstruction failures.   In this way it is determined that the simulated
efficiency for muon reconstruction requires a correction of $-4.4\%$ ($-1.1\%$) for muons with less than (greater than)
3 GeV/c.

\subsection{Pre-selections; calculation of neutrino energy} 

Although the two data sets involve different beams and final-state particles, 
they have many features in common.    In each case, 
reconstruction of CC event candidates proceeds by finding a long track that 
traverses both \minerva and MINOS and treating it as the muon.  The algorithm then 
searches for additional tracks that share a vertex 
with the longest track.  Kinked tracks, which are usually the 
result of secondary interactions, are then reconstructed by
searching for additional tracks starting at the endpoints of tracks previously found.
The differences in acceptance between $\mu^-$ and $\mu^+$ for the two samples are minor, and
backgrounds from wrong-sign muons are insignificant in either sample.  
Since the signal reactions are different, there are of course differences in selection cuts,
particle identification, and background subtraction procedures.  However after the 
final sample is obtained with each data set, the analyses the same method to extract the cross section.

To be accepted as a candidate event, a muon neutrino (antineutrino) interaction must have a $\mu^-$ ($\mu^+$) track
and the muon must originate within the central tracking volume.
The latter vertex is the primary interaction vertex;  it is required to be the only interaction
vertex within its time slice.   Furthermore the interaction vertex must lie within the fiducial volume.  
For the neutrino (antineutrino) events analyzed here, their primary vertices must
occur within the central 112 planes of the scintillator tracking region and must be
at least 20.5\,cm (22\,cm) from any edge of the planes.  
The fiducial volume contains a target mass of 5.57 (5.37) metric tons
with 3.54 (3.41) $\times 10^{30}$ nucleons.

The final-state muon momentum $p_{\mu}$ (and hence its energy $E_{\mu}$) is reconstructed
using the muon track's curvature or range measured in MINOS, in conjunction with the
track's $dE/dx$ energy loss from its observed traversal of the \minerva spectrometer.
The total final-state hadronic energy, $E_{H}$, is measured via calorimetry using the scintillator
light outputs generated by the final-state hadron shower particles.
More specifically, $E_{H}$ is obtained by scaling the calorimetric energy visible in the detector
according to the Monte Carlo detector response.
The neutrino energy $E_{\nu}$ is then calculated as follows:
\begin{equation}
\label{def-1}
  E_{\nu} = E_{\mu} + E_{H}. 
\end{equation}
\noindent
The resolution for $E_{\nu}$ determined in this way is 6\%. 
Further analysis procedures for the two samples
 -- their similarities and differences -- are described below.

\subsection{$\pi^{\pm}$ reconstruction;  $\numu$-CC($\pi^{+}$) selection}
\label{sec:pi+reco}

Events of the $\numu$-CC($\pi^{+}$) sample must have a reconstructed $\mu^{-}$ track that is
matched in the \minerva and \minos detectors, and the final state must have at least one
charged pion track.   Furthermore the reconstructed neutrino energy must be in the range
1.5\,--\,10\,GeV and the invariant hadronic mass must be less than 1.8\,GeV.      
There is no restriction on neutral pions, other mesons, or
baryons. Charged-current coherent pion production is included in the signal definition. 
In practice, only $\sim\, 5\%$ of selected events
have more than one charged pion.

Charged particle tracks are reconstructed by applying two pattern recognition algorithms 
to the clusters found within the tracking volume and downstream calorimeters. 
Charged pion tracks are identified using a containment requirement 
plus two particle-identification selections~\cite{Brandon-pion}. 
This results in an estimation of particle type and a determination of 
kinetic energy, $T_{\pi}$.

 A pion track is required to begin at the event vertex and to stop in either the tracking or electromagnetic-calorimeter regions 
of the central tracker.   This requirement restricts the maximum pion kinetic energy to 350\,MeV.  
The track is required to satisfy a particle-identification algorithm that evaluates the energy deposition pattern using
the Bethe-Bloch formula and -- very importantly -- to have, in the vicinity of its endpoint, 
a candidate Michel electron from the $\pi^{+}\rightarrow\mu^{+}\rightarrow e^{+}$ decay sequence~\cite{Brandon-pion}.  

The Michel selection disfavors negatively-charged pions that
tend to be captured on a nucleus before 
decaying, and discriminates strongly against pions 
that undergo charge exchange or absorption, thereby improving the pion energy resolution.
The efficiency for finding pion tracks 
with $T_{\pi}>50$\,MeV in simulated CC$(N\pi^{\pm})$ events (N\,=\,1,2) with $W<1.8$\,GeV, is 42\%.  The primary 
reasons for pion tracking inefficiency are secondary interactions of the pion
in the detector and activity in high-multiplicity events that obscures the pion.  

The invariant mass cut is on the experimentally-determined hadronic invariant mass, $W_{exp}$.
The cut was chosen to enrich the sample in events coming from
baryon resonances; in addition, the reconstruction efficiency is
higher because the final states have lower multiplicity.
Singly-produced $\pi^-$ tracks can only arise from FSI processes, and in any case 
the requirement that pion tracks of selected events have Michel electrons eliminates
most of them.   Consequently the selected pions are predicted to be 
$\pi^+$  at the level of 98.6\%.  The overall efficiency with which signal events having 
charged pions between 35\,MeV and 350\,MeV are selected is calculated to be 3\%.    
This value reflects reductions incurred as the result of 
the MINOS-matched muon requirement, the pion track reconstruction inefficiency (42\%), 
and the Michel electron selection.   According to the GENIE-based Monte Carlo simulation for this analysis,
the selected event sample has a signal purity of 86\%.

\begin{figure}
\begin{center}
\includegraphics[width=8.0cm]{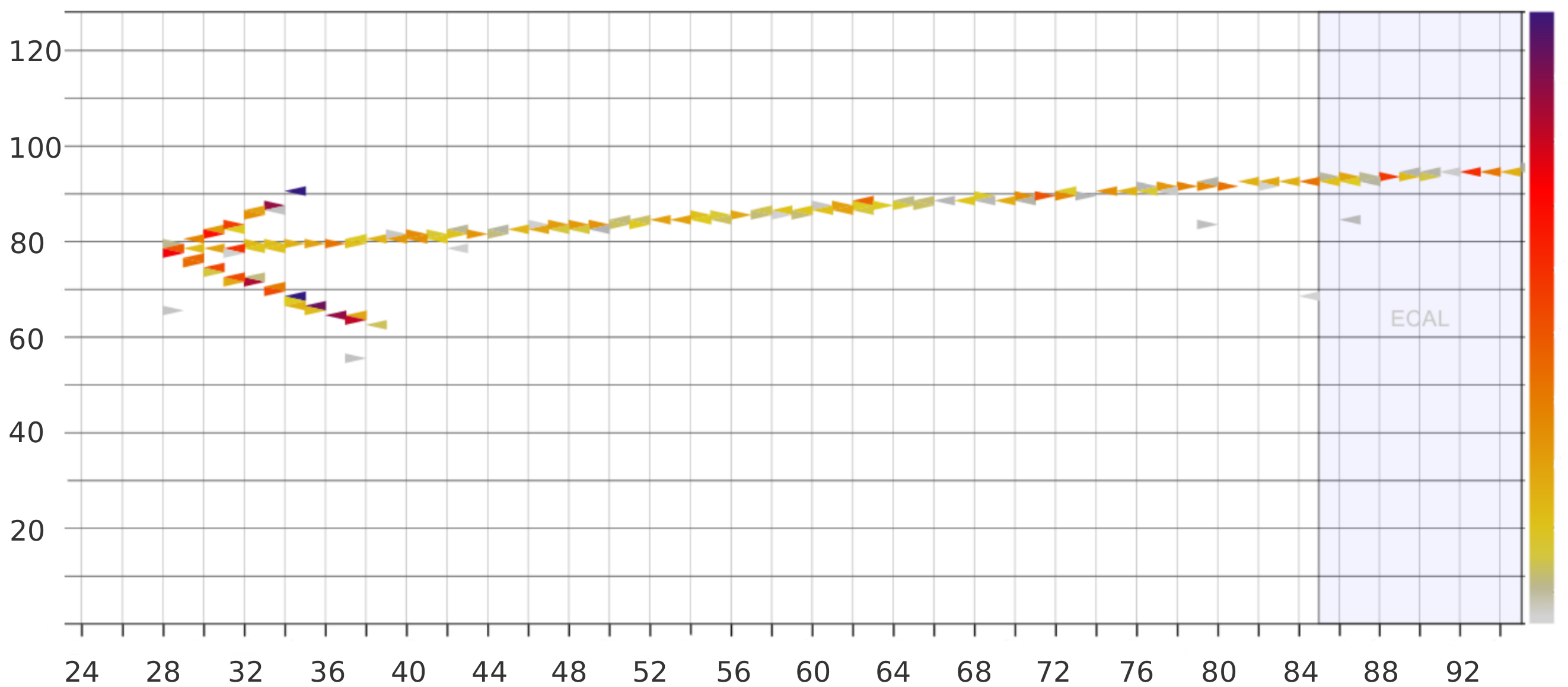}
\caption{Neutrino data event of the $\numu$-CC($\pi^{+}$) sample 
from the top view. The neutrino enters from the left.
The primary vertex occurs in the central scintillator tracking region; the muon traverses 
the downstream ECAL and HCAL regions and projects into the MINOS near
detector (downstream, not shown).  The event is a candidate for the final state 
$\mu^{-} \pi^{+} p$.  The X and Y axis labels show the module and strip numbers.   
The color (online) linear scale (0\,--\,10\,MeV) indicates 
the amount of energy deposited in the strips.
} 
\label{Fig01}
\end{center}
\end{figure}

Figure~\ref{Fig01} shows a $\numu$~+~hydrocarbon data event 
from the $\numu$-CC($\pi^{+}$) sample.   Emerging from the primary vertex in the 
central, plastic-scintillator tracking region are a muon track that exits downstream, 
a charged pion, and a short, heavily ionizing proton.   The projected image shows
vertical and horizontal spans in the detector medium of approximately 2.0\,m and 3.2\,m respectively.

\subsection{Selections for $\anumu$-CC($\pi^{0}$)}

For the $\anumu$-CC($\pi^{0}$) sample, the events must  
have a reconstructed $\mu^{+}$ track matched in the \minerva and 
\minos detectors, and the final state must contain a single $\pi^0$ 
unaccompanied by other mesons, with no restriction on the number of 
nucleons.  

\begin{figure}
\begin{center}
\includegraphics[width=8.0cm]{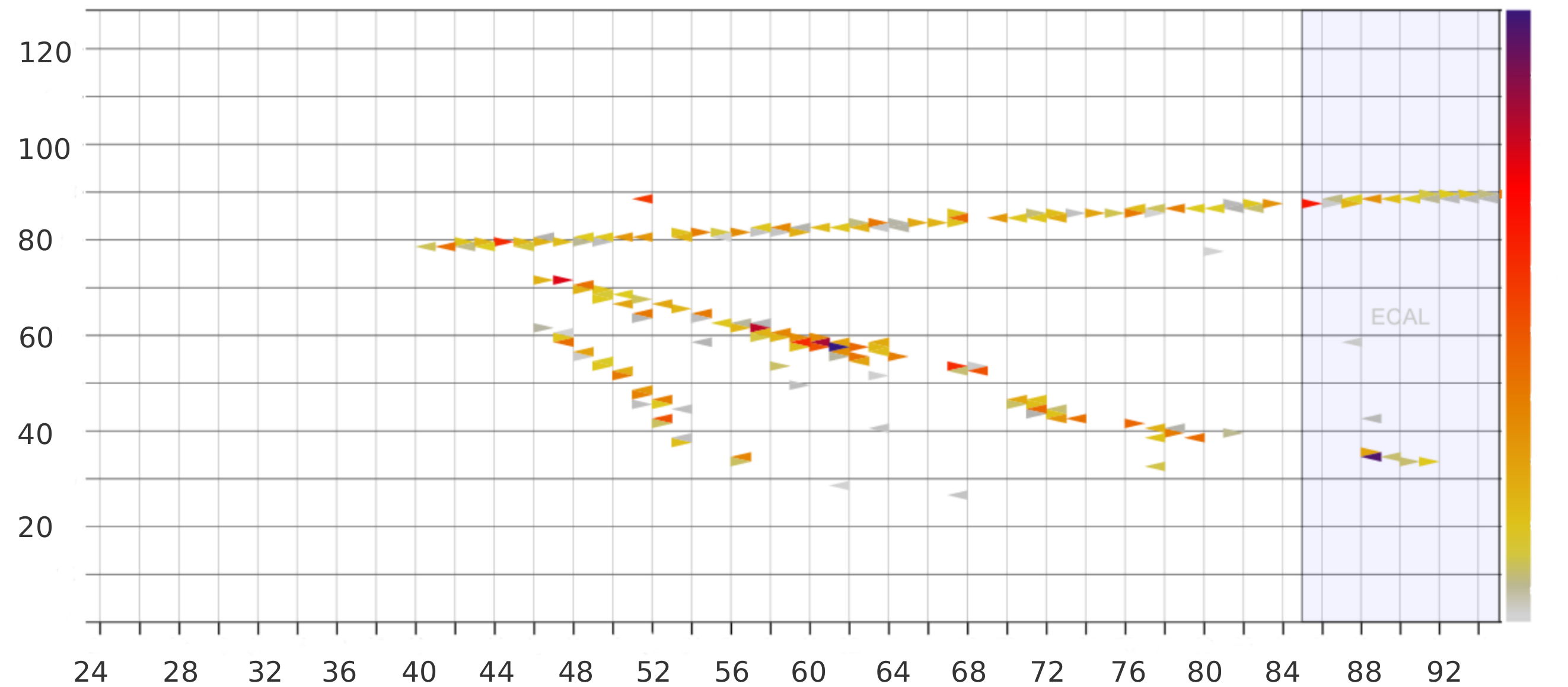}
\caption{Antineutrino data event and candidate for the final state $\mu^{+} \pi^{0} n$ from the top view. 
The event is initiated in the central scintillator tracker and extends into the downstream electromagnetic calorimeter.
It contains a $\mu^{+}$ track accompanied by two photon showers.} 
\label{Fig02}
\end{center}
\end{figure}

Candidate events contain 
a muon track in time coincidence with two electromagnetic showers.
Tracks that start within $\unit[5]{cm}$ of the vertex 
are considered to come from the primary vertex. 
Events that have primary tracks other than the muon track are rejected.    
Also removed are events that have isolated tracks
that do not point back to the vertex.  
Quite often, photons from the $\pi^0$ decays are reconstructed as tracks. 
Therefore, tracks with separation distance 
greater than $\unit[5]{cm}$ from and pointing back to the event vertex are not considered as
coming from the vertex, and their associated energy clusters are made available to the $\pi^0$ reconstruction.
The total visible energy in the tracker, 
electromagnetic calorimeter (ECAL), and hadronic calorimeter 
(HCAL) is required to be greater than 80 MeV and less than 2 GeV. 
The cut on low visible energy removes events whose 
total energy deposition is too low to encompass the $\pi^0$ rest mass. 
The upper visible energy cut removes deep inelastic scattering (DIS) background events.
Candidate events are allowed to have
isolated hit clusters in the vicinity of the vertex because these may be 
induced by final-state nucleons interacting with the
hydrogen or carbon nuclei of the detector. 
The data event shown in Fig.~\ref{Fig02} 
is a $\anumu$ + hydrocarbon interaction that exhibits the
prerequisite properties for retention in the $\anumu$-CC($\pi^{0}$) candidate sample.

Events that satisfy the topological selections 
are passed to the $\pi^0$ reconstruction.  The reconstruction proceeds in two stages.
In the initial stage pattern recognition is performed to identify the two gamma showers.
Hit clusters found in the X view that are close in polar angle with respect to
the vertex, but can be separated in radial distance from the vertex, 
are grouped into photon-conversion candidates. 
Then, for each candidate, clusters in the U and V views 
that are consistent among the three views are added. 
Photon candidates must have clusters in at least two views 
in order to enable their directions to be reconstructed in three dimensions.   

Reconstruction of the photon showers is carried out in the second stage.   The
position, direction, and energy of a photon shower are determined by
the clusters that have been assigned to each of the candidate photons. 
The photon direction is reconstructed from the cluster energy-weighted slopes in each view.
The photon vertex is defined using the closest cluster to the event
vertex on the photon direction axis. 
The photon energy is reconstructed by calorimetry using calibration constants 
determined by detector response simulations. 

Candidate events must have exactly two reconstructed photon showers.
In order to reduce charged-pion backgrounds,
each photon is required to have converted at least $\unit[15]{cm}$ ($0.36$ radiation length) 
away from the primary vertex. 
The two-photon invariant mass $m_{\gamma\gamma}$ is reconstructed 
from the photon energies $E_1,E_2$ and
the separation angle $\theta_{\gamma\gamma}$ between the two photons using
\begin{equation}
m_{\gamma\gamma}^2=2E_1E_2(1-\cos\theta_{\gamma\gamma}). 
\label{eq:invmass}
\end{equation}
The overall calibration constant that sets the absolute energy scale is
determined by matching the peak in the $\gamma\gamma$ invariant mass
distribution to the nominal $\pi^0$ mass. 
This procedure is done separately for data and simulation which enables
correction for a difference in energy scales of 5\% with 2.2\% uncertainty between the data and simulation. 
Finally, the $\pi^0$ momentum is calculated from momentum conservation, $\vec p_{\pi^0}
= \vec{k}_1 + \vec{k}_2$, where $\vec{k}_i$ are reconstructed photon momenta. 
The $\pi^0$ reconstruction typically has a 25\% energy resolution and 3.5$^{\text o}$ angular resolution in each view. 

It is required that the invariant mass $m_{\gamma\gamma}$ lies between
$\unit[75]{MeV/c^2}$ and $\unit[195]{MeV/c^2}$, and that $E_\nu$ falls 
between 1.5 and 10 GeV.  The lower cut on $E_\nu$ maximizes 
MINOS acceptance while the upper cut reduces flux uncertainties. Additionally, 
the reconstructed $W$ is limited to $W <\,\unit[1.8]{GeV}$.

\subsection{Reconstruction of $Q^{2}$ and $W$}
\label{sec:q2w}
Calculation of the four-momentum-transfer-squared, 
$Q^{2}$, and of the hadronic invariant mass, $W$, proceeds according to
\begin{equation}
\label{def-2}
Q^2 = -(k-k')^2 =  2E_\nu(E_\mu-|\vec{p}_\mu|\cos\theta_{\mu})-m_\mu^2 ,
\end{equation}
and
\begin{equation}
\label{def-3}
 W^2 = (p+q)^2 = M_N^2 + 2M_N(E_\nu-E_\mu)- Q^2 ,
\end{equation}
where $p$ is the four-momentum vector of the initial nucleon,
$q=k-k'$ is the four-momentum transfer, and $M_N$ is the nucleon mass. 

The calculations for $E_\nu$ (Eq.~\eqref{def-1}) and $Q^2$ (Eq.~\eqref{def-2})
do not involve any assumption concerning the state of the initial nucleon 
momentum or the composition of particles in the final state.
On the other hand, the prescription for estimation of $W$ in Eq.~\eqref{def-3}
assumes an initial-state nucleon at rest.   The Monte Carlo uses a relativistic
global Fermi Gas nuclear model; this model is known to be accurate
for high momentum transfers (roughly $Q^2 > 1$\,GeV$^2$) but less accurate 
for low momentum transfers.  The rms widths of the $Q^2$ and $W$ 
variables are 18\% and 8\%, respectively, for the charged
pion analysis and 16\% and 10\% for the neutral pion analysis.

\section{Determination of cross sections}

After particle identification, the $\numu$-CC($\pi^{+}$) sample contains 5410 events.  
For the full sample of signal events, the efficiency is 1.25\% and the purity is 86\%.  
The total background is estimated using the distribution of the
reconstructed invariant mass, $W_{exp}$, shown in Fig.~\ref{Fig03}.
The largest contribution to the background (69\%) is estimated 
to arise from pion production at true (simulation) hadronic invariant mass values, $W_{true} $,
greater than 1.8\,GeV.   Events with protons misidentified as pions account
for 19\% of background, while events with $E_{\nu} > 10$\,GeV, primary vertices outside the fiducial
volume, and neutral current events account for the remaining 9\%, 2\%, and 1\% respectively.

The selected $\anumu$-CC($\pi^{0}$) sample contains 1004 events. 
The total selection efficiency is 6\% and the purity is 55\%.    
It is estimated that 70\% of the total background is populated
by antineutrino interactions that produce at least one $\pi^0$ in the detector. 
The background is nearly equally comprised of multi-pion production events, 
e.g. $\pi^0+\pi^{\pm}$, where the $\pi^{\pm}$ is not tracked, and
events with a secondary $\pi^0$ produced by $\pi^-\rightarrow\pi^0$ 
charge exchange or by nucleon scattering in the detector volume.  
The remaining 30\% of background events are non-$\pi^0$ events wherein 
$\pi^-$ and neutron-induced ionizations are mistakenly identified as photons.

\subsection{Background subtraction and unfolding} 
\label{sec:wbkgd}
Cuts are made in both analyses to focus on the response in the
kinematic region dominated by baryon resonances ($W <$ 1.8\,GeV).
However the $W_{exp}$ cut described in Sect.~\ref{sec:pi+reco} is
insufficient to get the desired measurement. 
A significant background comes from true pion production 
at higher $W$; for the selected charged pion sample this background
comprises 6\% of the sample.   To remove this background,
events are first selected with $W_{exp}<$ 1.8\,GeV with $W$ 
calculated as in Sect.~\ref{sec:q2w}.  Then, background is subtracted 
according to $W_{true}$ (the value of $W$ at the primary interaction 
according to the Monte Carlo) through a sideband procedure~\cite{Brandon-thesis}.
The $\anumu$-CC($\piz$) analysis includes an additional background subtraction 
based upon the $m_{\gamma \gamma}$ spectrum.
 
\begin{figure}
  \begin{center}
                 {
                \includegraphics[width=8.5cm]{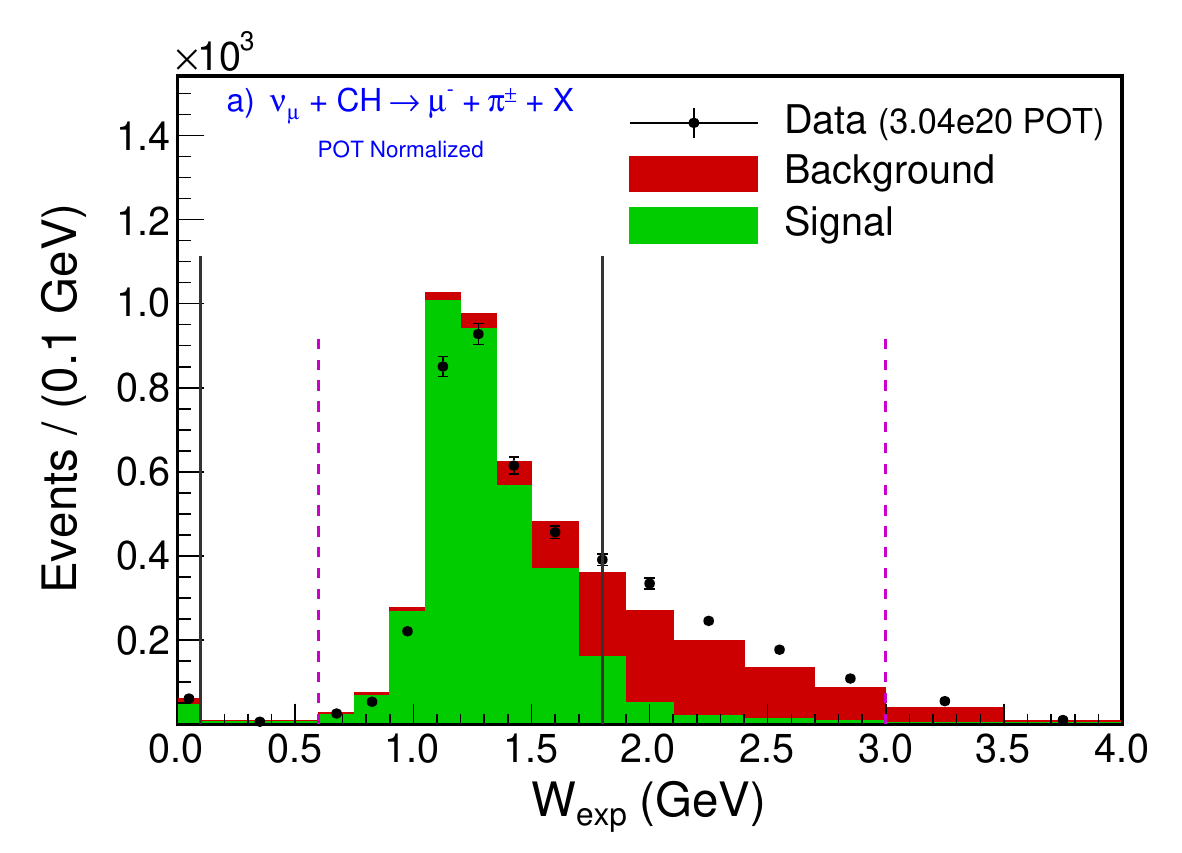}
                             }
              
                        {
                          \includegraphics[width=8.5cm]{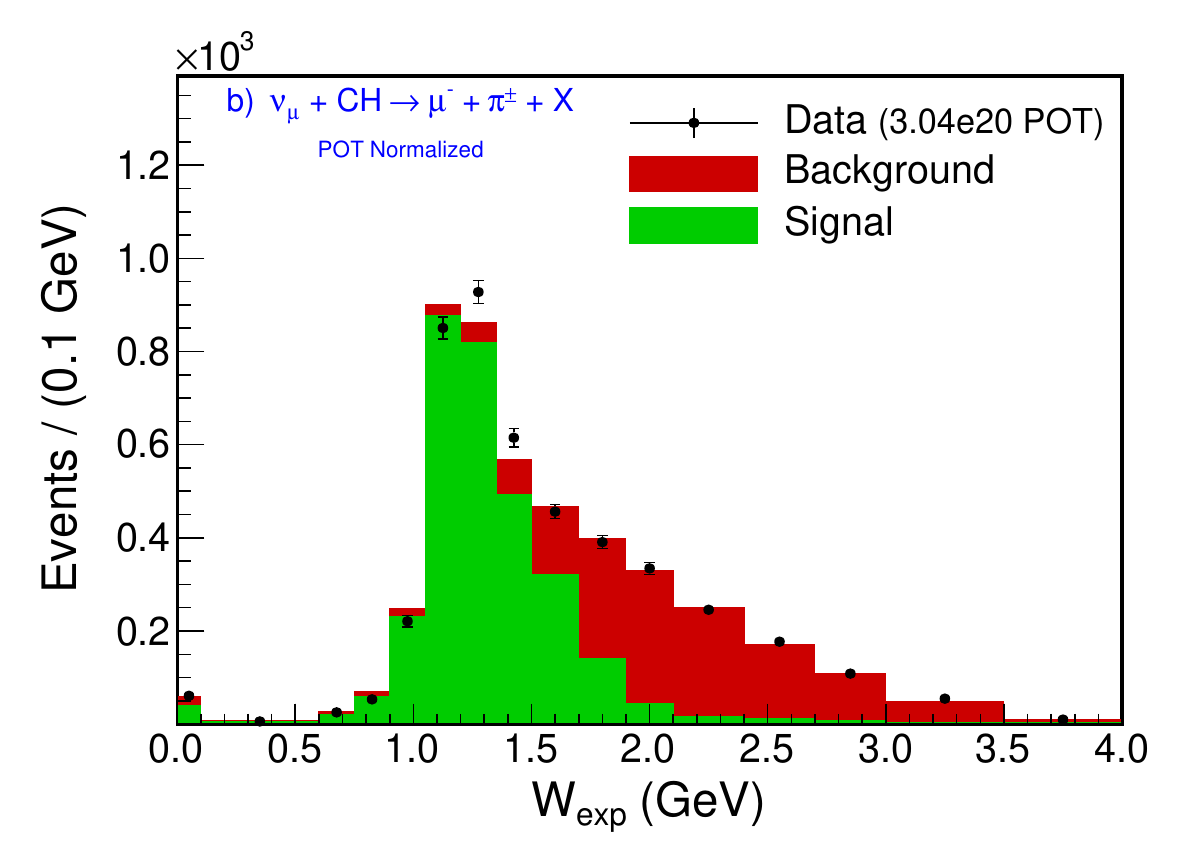}
                                         }
\caption{The $W_{exp}$ spectrum for the $\numu$-CC($\pi^{+})$ sample
comparing the data (solid points) to the Monte Carlo simulation 
(histograms) before (a) and after (b) the fit over the region between 
the vertical dashed lines in the upper plot.  The Monte Carlo prediction 
is the sum of signal template (green online) and background template 
(red online) contributions.  The analysis selects a signal-dominated
sample (between the vertical solid lines) by requiring $W_{exp}<1.8$ GeV.}
\label{Fig03}
\end{center}
\end{figure}

 For the background subtraction based upon $W_{true}$, 
the simulated $W_{exp}$ distribution is 
divided into signal and background templates according to 
$W_{true}<\unit[1.8]{GeV}$ (signal) and $W_{true}>\unit[1.8]{GeV}$ 
(background) as indicated in Fig.~\ref{Fig03}a.
The templates are then fitted, bin-by-bin, to the $W_{exp}$ spectrum;
the normalizations of the signal and background 
templates are the fit parameters in maximum likelihood fits (see Ref.~\cite{Brandon-thesis} for details).  
The full $W_{exp}$ spectrum after the fit 
is shown in Fig.~\ref{Fig03}b.  

\begin{figure}
  \begin{center}
     \includegraphics[width=8.5cm]{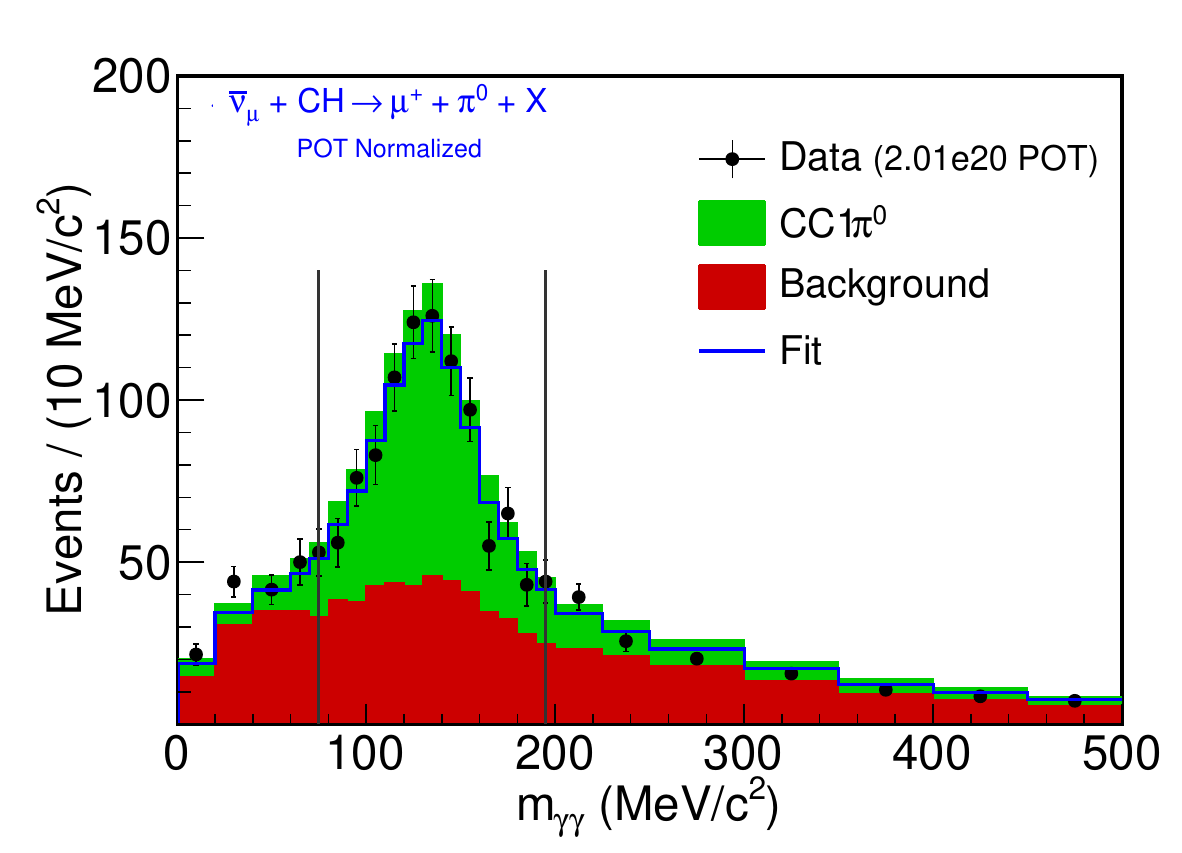}
\caption{Distribution of the invariant mass of the $\gamma\gamma$ pair. Data
are shown as solid circles with statistical error bars. 
The shaded histograms show the Monte Carlo predictions for the $\anumu$-CC($\piz$) signal 
and background. The solid blue (online) histogram is the maximum-likelihood best fit to the data.
The vertical lines indicate the invariant mass cut, 
$\unit[75]{MeV/c^2} < m_{\gamma\gamma} < \unit[195]{MeV/c^2}$, that defines the signal region.
}
\label{Fig04}
\end{center}
\end{figure}

For the $\anumu$-CC($\piz$) sample, the background is constrained using the two-photon
invariant mass $m_{\gamma\gamma}$ distribution.
Figure~\ref{Fig04} shows the $m_{\gamma\gamma}$ distribution of the data and shows
the contributions from signal and background that are estimated by the MC simulation.
The data $m_{\gamma\gamma}$ distribution is fitted to a $m_{\gamma\gamma}$ model using 
the binned extended maximum-likelihood method. The model is constructed
from the shapes of the MC signal and background event distributions. The expected numbers of signal and
background events are parameters determined from the fit.
The result of the fit is shown as solid histogram in the same figure. 
The fit reduces the background normalization in the signal region by 11\% 
compared to the simulation prediction.

Resolution effects in the scintillator are simulated with Monte Carlo and are 
unfolded from the data using a Bayesian procedure~\cite{D'Agostini:1994zf}.  
For either event sample, the unfolding matrices are close to diagonal and
so the effects of unfolding are minor.

\subsection{Cross section calculation}
\label{X-sec-calc}

As in Refs.~\cite{Brandon-pion,Trung-pion}, the 
flux-integrated differential cross section per nucleon
for kinematic variable $X$ (such as  $\theta_\mu$, $p_{\mu}$, and $Q^{2}$), 
in bins of $i$, is calculated according to
\begin{equation}
\label{eq:dif-xsec}
\left( \frac{d\sigma}{dX} \right)_{i} =  \frac{1}{T\Phi } \frac{1}{\Delta X_i} 
	\frac{\sum\limits_{j} U_{ij} \left( N^{data}_{j} - N^{bkg}_{j} \right) }{\epsilon_{i}}.
\end{equation}
Here, $T$ is the number of nucleons in the fiducial volume,
$\Phi$ is the integrated flux, $\Delta X_i$ is the bin width,
$\epsilon_{i}$ is the selection efficiency and acceptance.   The  
unfolding function, $U_{ij}$, calculates the contribution
to true bin $i$ from reconstructed bin $j$, with the 
number of data candidates, $N_{j}^{data}$, and the number of 
estimated background events, $N_{j}^{bkg}$.  Both the efficiency
$\epsilon$ and the unfolding matrix $U$ are estimated using the simulation.
The total cross section in neutrino energy is calculated in a slightly different way:
\begin{equation}
  \sigma(E_\nu)_i =  \frac{1}{T\Phi_i}  
	\frac{\sum\limits_{j} U_{ij} \left( N^{data}_{j} - N^{bkg}_{j} \right) }{\epsilon_{i}}. 
\end{equation}
Here, the total flux $\Phi_i$ is calculated for each bin of neutrino 
energy.  Both the integrated and binned fluxes are calculated by Monte Carlo methods which
take into account the full geometry of the production target region
and all processes by which pions and kaons are produced and 
subsequently decay.  

The present analysis benefits from updated 
flux calculations~\cite{minerva-flux-2015}.   In addition, a flux constraint 
provided by the \minerva $\nu + e^{-}$ scattering measurement~\cite{Park-thesis} has been applied.
The constraint derived from measurement of muon-neutrino elastic scattering on electrons yielded 
a fractional change of between 1-2\% in absolute rate for the range 1.5 GeV $< E_{\nu} < 10$ GeV.   
Larger changes arose from revision of the absolute fluxes downward by 11-12\% 
upon constraining them to hadron production data and incorporating improved determinations 
of the beamline geometry~\cite{minerva-flux-2015}.   Together, these changes resulted 
in an upward shift of absolute event rate for the 
neutrino (antineutrino) exposure of 13\%\,(12\%) 
averaged over the analyzed $E_{\nu}$ range.
In each event sample, changes to energy dependence introduced by
 the updated fluxes are small;  the changes are almost entirely
in absolute magnitude for all results except for the neutrino energy cross section.

The revisions to pion production differential cross sections 
reported in~\cite{Brandon-pion,Trung-pion} are sizable; 
the updated cross sections are given in the Appendix.   
The revised values fall beyond the 1$\sigma$ flux uncertainties of 9\% and 10\% 
indicated by the earlier works, however they remain within the 
overall systematics uncertainty envelopes given in those works.

\section{Systematic Uncertainties}
Systematic uncertainties are evaluated in almost identical ways
for the two event samples analyzed here.  The methods used are described
in Ref.~\cite{Brandon-pion}.  The systematic uncertainty 
from the neutrino flux is described in detail in Refs.~\cite{minerva-flux-2015, nubarprl}.  
Although the uncertainties arise from many individual sources, 
they can be grouped into five categories as being associated 
with the detector energy response ${\it (i)}$,  
with the principal-process models ${\it (ii)}$ and final-state interaction models 
used by the reference Monte Carlo (GENIE) ${\it (iii)}$, 
and with the neutrino flux ${\it (iv)}$.  Among the 
remaining odd-lot of sources designated as ``other" ${\it (v)}$,
subtraction of background gives the largest uncertainty in either data set.

\subsection{$\pi^+$ production}
Cross-section uncertainties are shown for muon momenta of the 
$\numu$-CC($\pi^{+}$) sample in Fig.~\ref{Fig05}a.
Uncertainty from systematic sources slightly exceeds the statistical 
error for this sample.  No single source dominates the systematic error.  
The largest contribution arises from the detector energy response, 
which is expected because the hadronic energy is measured by the energy 
deposited in scintillator layers.    The uncertainty in the neutrino flux is smaller than was the 
case for Refs.~\cite{Brandon-pion,Trung-pion}, but it is still the second-largest source of uncertainty.
The uncertainty associated with neutrino cross sections is somewhat smaller than the flux uncertainty.

\subsection{$\pi^0$ production}

\begin{figure}
  \begin{center}
              {
                \includegraphics[width=8.5cm]{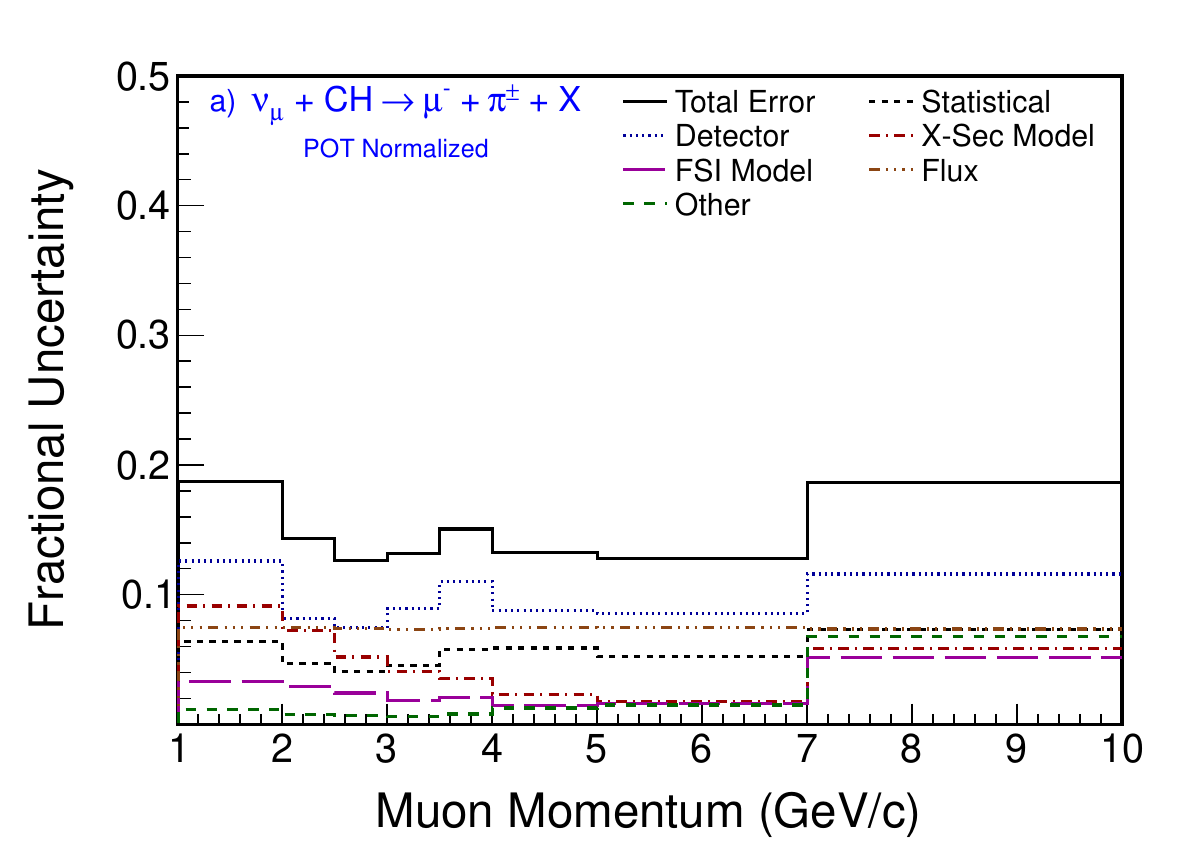}
              }
                        {
                          \includegraphics[width=8.5cm]{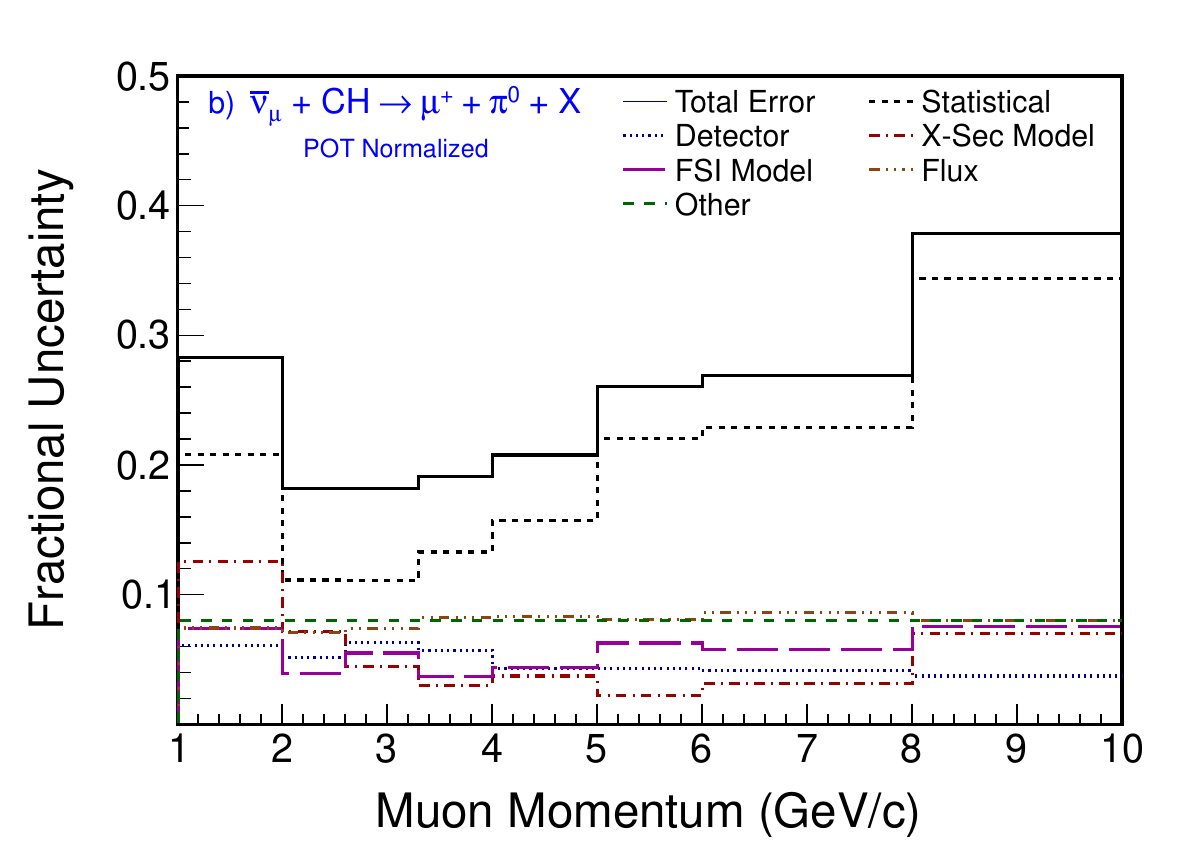}
                        }
\caption{Fractional uncertainties for muon momenta in the $\numu$-CC($\pi^{+}$)
analysis (a), and the $\anumu$-CC($\pi^{0}$) analysis (b), for the 
case of absolute normalization to the data exposure.  
The statistical (total) error is shown by the dashed (highest solid-line) histogram 
in each plot.   Component histograms show 
the contributions from the five systematic error sources.}
\label{Fig05}
\end{center}
\end{figure}

Figure~\ref{Fig05}b shows the cross-section uncertainties 
for muon momenta of the $\anumu$-CC($\pi^{0}$) sample.
For this sample, the the statistical uncertainty of the limited data set (dashed histogram)
is larger than either the flux or cross-section uncertainties. 
The normalization of the background fit contributes 8\% to the systematic uncertainty.
Significant uncertainty arises from the misidentification of neutrons as photons.
This source of error was evaluated by changing the neutron inelastic
cross section within an error range based upon compiled measurements.
A large fraction of the secondary $\pi^0$ in the background is estimated to arise from $\pi^{-} \rightarrow \pi^0$
charge exchange (CEX), for which the cross sections are poorly known. The effect of this uncertainty on our 
measurement is evaluated by changing the CEX cross section within its 
uncertainty of $\pm$50\%~\cite{Ashery:1981tq,Bowles:1981,Jones:1993}, and then 
re-measuring the cross sections.  The uncertainty in the 
electromagnetic energy scale contributes 2.2\% to the error budget, estimated
from the fitted mean uncertainty of the data $m_{\gamma\gamma}$ distribution.

The principal-interaction cross-section model (GENIE) also contributes significantly to the uncertainty of both analyses.
One of the large uncertainties arises from modeling the basic pion production process on individual nucleons.  
Tables of values of the cross section and of systematic uncertainty decomposition 
for each bin of each measurement are given in the Supplement to this paper~\cite{Supplement}.

\section{Muon Kinematics in CC($\pi$) Production}

\subsection{Muon production angle}
\begin{figure}
  \begin{center}
              {
                \includegraphics[width=8.5cm]{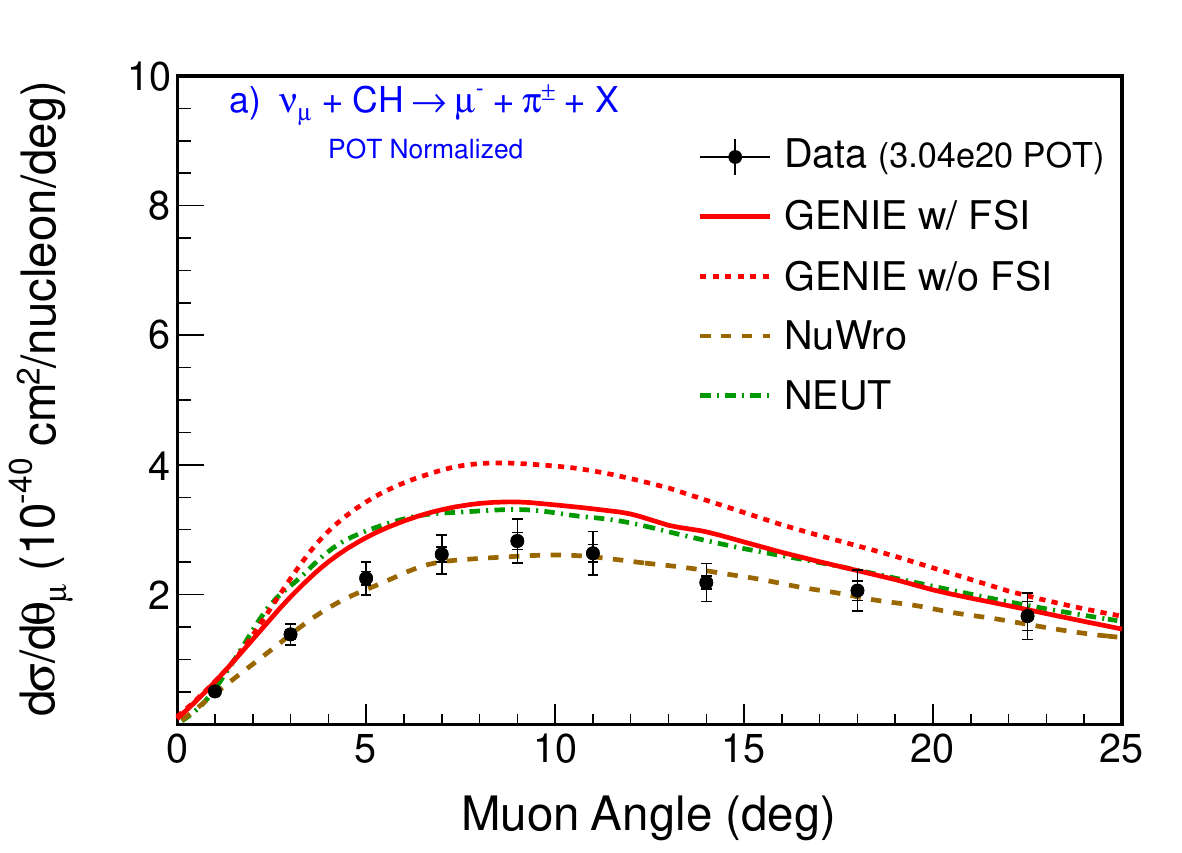}
              }
                        {
                          \includegraphics[width=8.5cm]{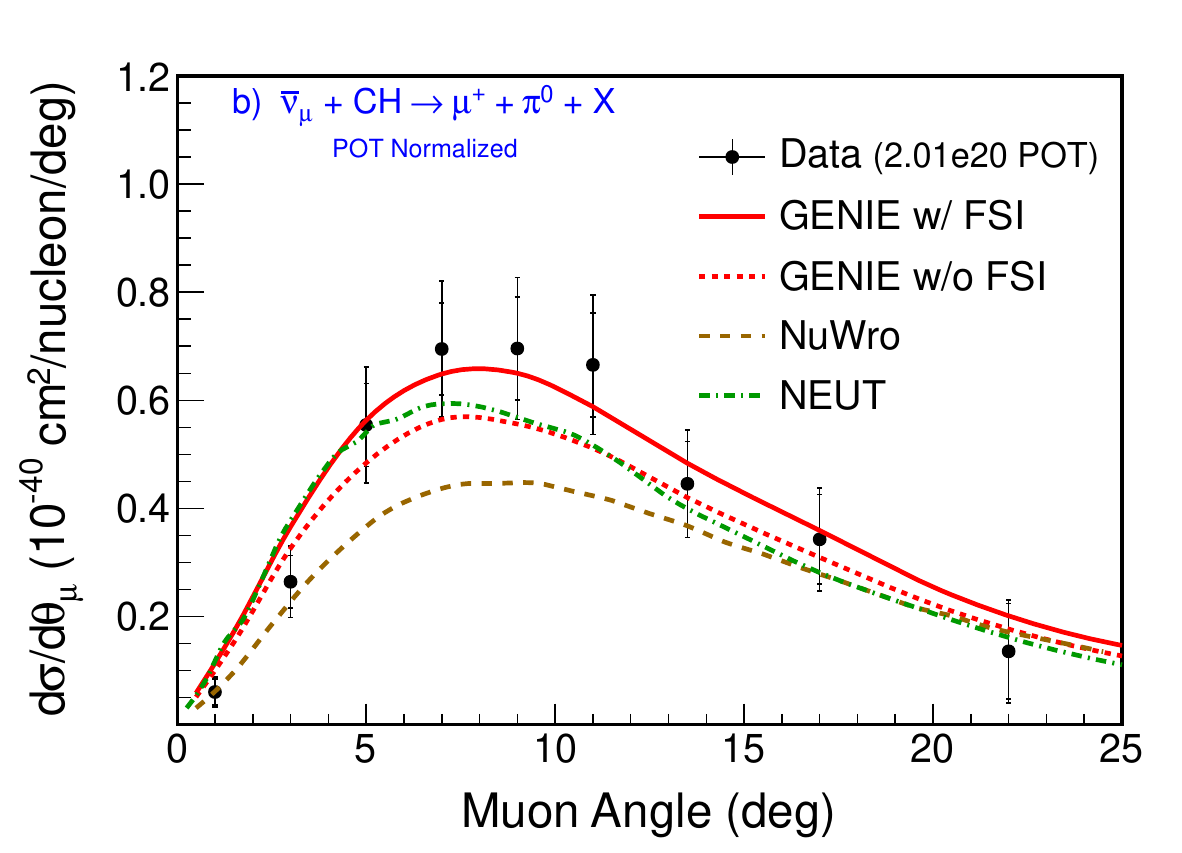}
                        }
\caption{Differential cross sections as a function of the muon production angle $\theta_\mu$  
	for the $\numu$ (a) versus $\anumu$ (b) pion production samples.
	Data are shown as solid circles. The inner (outer) error bars correspond to statistical (total) uncertainties.
        The solid (dashed) distributions are GENIE predictions with (without) FSI.   
        Predictions for the NuWro and NEUT event generators are also shown.}
\label{Fig06}
\end{center}
\end{figure}

Figure~\ref{Fig06} shows the differential cross sections 
as a function of polar angle, $\theta_\mu$, with respect to the 
neutrino beam,  for the $\numu$-CC($\pi^{+}$) sample 
(Fig.~\ref{Fig06}a) and for the $\anumu$-CC($\pi^{0}$) sample (Fig.~\ref{Fig06}b). 
For both samples, the $\theta_\mu$ distribution peaks around $8^\circ$ and then decreases gradually.  Beyond $25^\circ$ the
acceptance into the MINOS near detector is small, and so no cross sections are given for that region.
The superimposed solid-line (dashed-line) histogram shows the GENIE prediction 
 that includes (omits) the intranuclear FSI treatment. 
The ratio of the predictions with/without FSI is observed to be roughly constant 
over the observed angular range.

Comparison of the dashed and solid-line histograms in 
Fig.~\ref{Fig06}a,b
shows that pion FSI play a significant role in the GENIE predictions.
 In the $\numu$-CC($\pi^{+}$) sample (upper plot), 
 $\Delta(1232)$ production in the charge state $\Delta^{++}$
dominates the final state, and pion intranuclear absorption plus pion 
charge exchange deplete the number of final-state pions that exit the nucleus.
This depletion cannot be compensated by charge-exchange feed-in 
from $\Delta^{+}$ channels which are produced at lower rates (due to their
smaller isospin amplitudes).    Thus, for reactions \eqref{reaction-1}, 
the GENIE prediction with FSI included is always smaller than
the GENIE prediction without FSI.    For reactions \eqref{reaction-2} 
of the $\anumu$-CC($\pi^{0}$) sample however, the situation is reversed.    
The latter reactions also lose pions to intranuclear absorption and charge exchange. 
However, the feed-in of charge-exchanged
$\pi^{0}$ originating from production of $\Delta^{-}$ states is always larger than the losses.  
Production of the latter states benefits from having a relatively large isospin amplitude.
The net result is that for reactions \eqref{reaction-2} the GENIE prediction is  
elevated by the inclusion of FSI processes (Fig.~\ref{Fig06}b).

Differences in absolute rate between the data and GENIE predictions are 
evident in Fig.~\ref{Fig06}a.
The GENIE prediction with FSI (solid-line curve) is too high by 20\% to 30\% for the 
neutrino-induced sample of Fig.~\ref{Fig06}a.   
On the other hand, Fig.~\ref{Fig06}b shows
the GENIE prediction with FSI to be in good agreement  
with the distribution for the antineutrino sample.
The uncertainties associated with the absolute $\numu$ and $\numubar$ fluxes
are 8.5\% and 8.0\% respectively, so the data/MC normalization differences are of order
$2.4\,\sigma$ and $ 0.3\,\sigma$.

In Fig.~\ref{Fig06} and in subsequent figures, predictions of 
the NEUT and NuWro neutrino event generators  are displayed with FSI effects included, 
providing comparisons with GENIE as well as additional predictions for the data.
For baryon-resonance production, NEUT (like GENIE)
 uses the Rein-Sehgal model~\cite{Rein:1980wg} 
 but without inclusion of baryon-resonance interference, 
 whereas NuWro includes only $\Delta(1232)$ production 
 as formulated by the Adler model~\cite{Adler:1968, Adler:1975}.    
 NEUT also incorporates nonresonant pion production 
 from Rein-Sehgal,  whereas NuWro (like GENIE) uses 
 the Bodek-Yang model~\cite{Bodek:2004pc} 
 above the resonance region and extrapolates it to lower $W$ 
 so as to converge with the predictions of Rein-Seghal.   
 For their FSI treatments, both NEUT and NuWro use 
 the Salcedo-Oset model~\cite{Salcedo-Oset} in a cascade formalism that includes nuclear medium corrections. 
The NEUT and NuWro predictions are compared 
to data for which the background estimates have been launched from predictions of the 
GENIE event generator.   The GENIE predictions however are constrained 
by data in the sidebands; moreover the full systematics uncertainties arising from GENIE as well as
other sources are taken into account in the predicted backgrounds.   Consequently
any biasing of measurements towards GENIE predictions will fall within the
systematics error envelope indicated for the data points.

\smallskip
Figure~\ref{Fig06} shows that all three event generators achieve good agreement for the shape 
of $d\sigma/d\theta_{\mu}$ for both of the CC pion production samples.
NEUT, like GENIE, predicts an absolute rate for the $\numu$-CC($\pi^{+})$ sample that is
distinctly higher than for the data, while the NuWro prediction for the same sample
is in excellent agreement with respect to distribution shape and normalization.
For the $\anumu$-CC($\pi^{0}$) sample of Fig.~\ref{Fig06}b,
however, the situation is opposite:    GENIE and NEUT achieve good agreement
in normalization as well as shape, while the NuWro prediction falls well below the data.

\begin{figure}
  \begin{center}
              {
                \includegraphics[width=8.5cm]{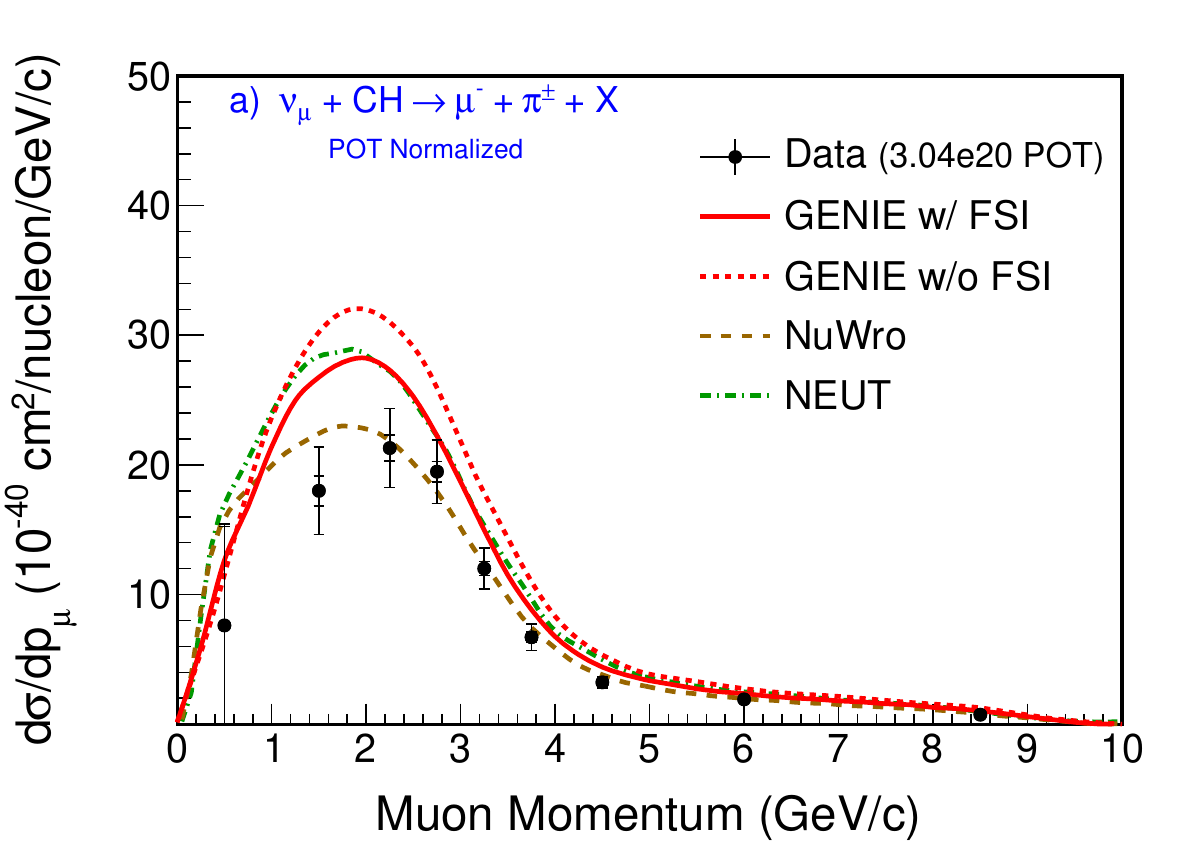}
              }
                        {
                          \includegraphics[width=8.5cm]{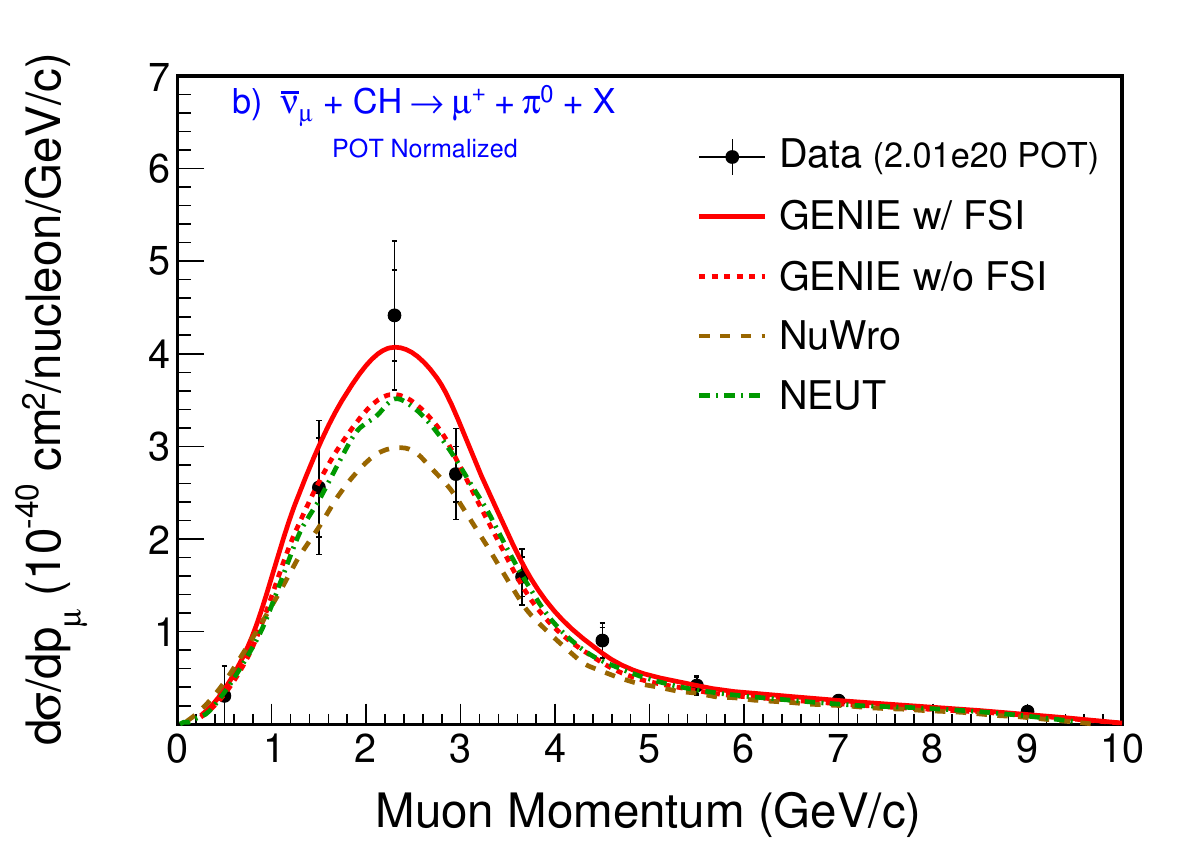}
                        }
\caption{Differential cross sections as a function of the muon momentum for the 
$\numu$ (a) and $\anumu$ (b) pion production samples. Data are shown as solid circles.  
The inner (outer) error bars correspond to statistical (total) uncertainties.
        The solid (dashed) distributions are GENIE predictions with (without) FSI.  
        Also shown are predictions for the NuWro and NEUT event generators.}
\label{Fig07}
\end{center}
\end{figure}

Normalization differences between data and predictions of event generators can be driven by cross-section uncertainties
for $\numu/\numubar$-induced pion production in scattering on free nucleons, as well as by flux uncertainties.
The data constraining these processes are sparse, 
and in the case of $\numu + p \rightarrow \mu^{-} + \pi^{+} + p$,  
the two bubble chamber measurements using 0.5 to $\sim3$ GeV neutrinos
reported cross sections that differed by $\sim 30\%$ in absolute normalization~\cite{radeckyanl,kitagakibnl}.
(A recent reanalysis obtains consistency between these two data sets~\cite{Wilkinson} , 
however the generator predictions shown here have not been tuned to the results of this reanalysis.)
For the channel $\numubar + p \rightarrow \mu^{+} + \pi^{0} + n$ there is only one cross-section data point, obtained
with antineutrino scattering on a heavy-liquid (freon CF$_3$Br) bubble chamber fill~\cite{skat}.

\subsection{Muon momentum}
The differential cross sections as a function of the muon momentum $p_\mu$ 
for the two samples are shown in Fig.~\ref{Fig07}.
The distributions peak between 2.0 and 2.5 GeV and fall off rapidly 
as $p_\mu$ increases from 3.0 to beyond 6.0 GeV.  
The same trends as observed in $d\sigma/d\theta_{\mu}$ are apparent 
here in $d\sigma/dp_{\mu}$.  The relatively large uncertainty for the lowest-momentum bin 
in Fig.~\ref{Fig07}a is an artifact of the muon 
acceptance in MINOS.  Because muons with $p_{\mu} < 1$~GeV have low 
efficiency, the data selection requires $E_\nu >$ 1.5 GeV.  Therefore, 
the first bin only receives event counts as the result of the 
unfolding procedure.

Figure~\ref{Fig07} compares the observed 
$d\sigma/dp_{\mu}$ distributions to predictions of the three event generators.
Similar to the situation with $d\sigma/d\theta_{\mu}$ 
in Fig.~\ref{Fig06}, the three generators achieve good agreement 
with respect to the shape of $d\sigma/dp_{\mu}$ for both samples, but with variance in the predictions for absolute rates.

\section{Composition of data samples}

\begin{figure}
  \begin{center}
              {
                \includegraphics[width=8.5cm]{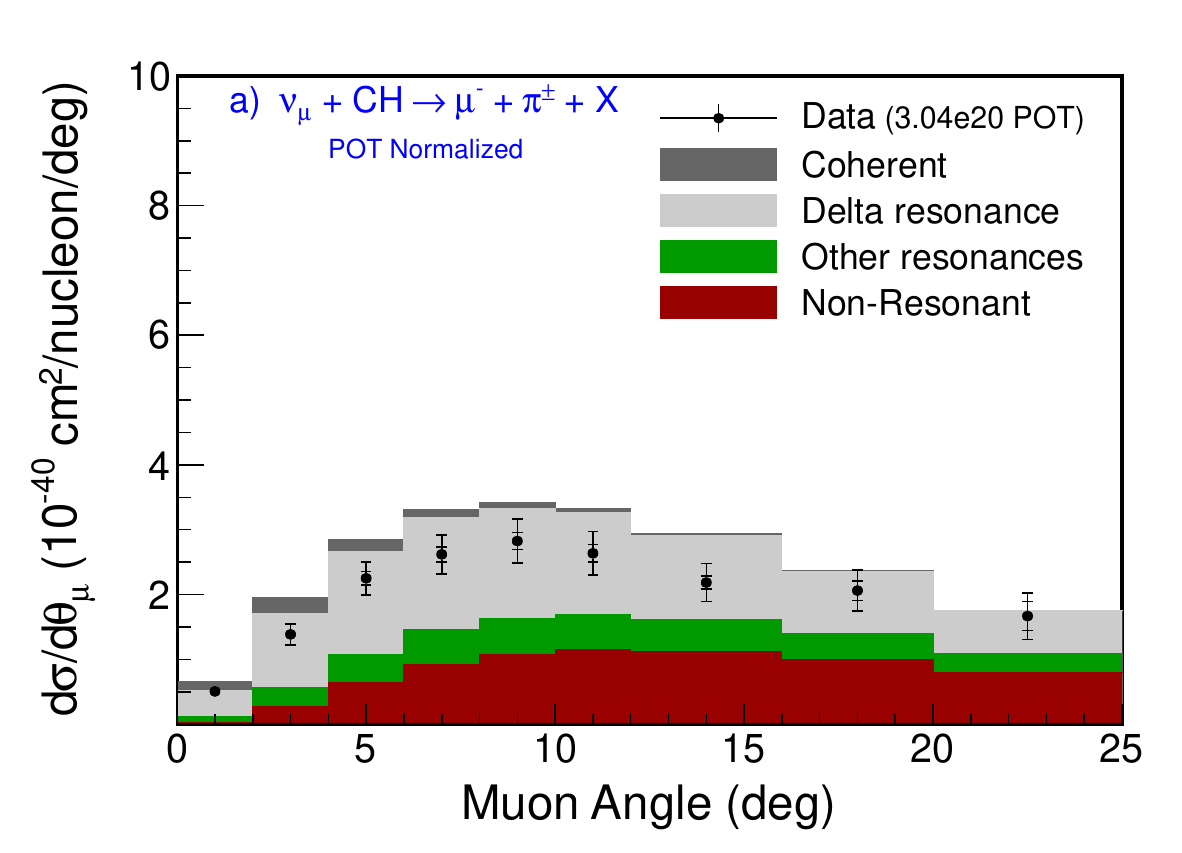}
              }
                        {
                          \includegraphics[width=8.5cm]{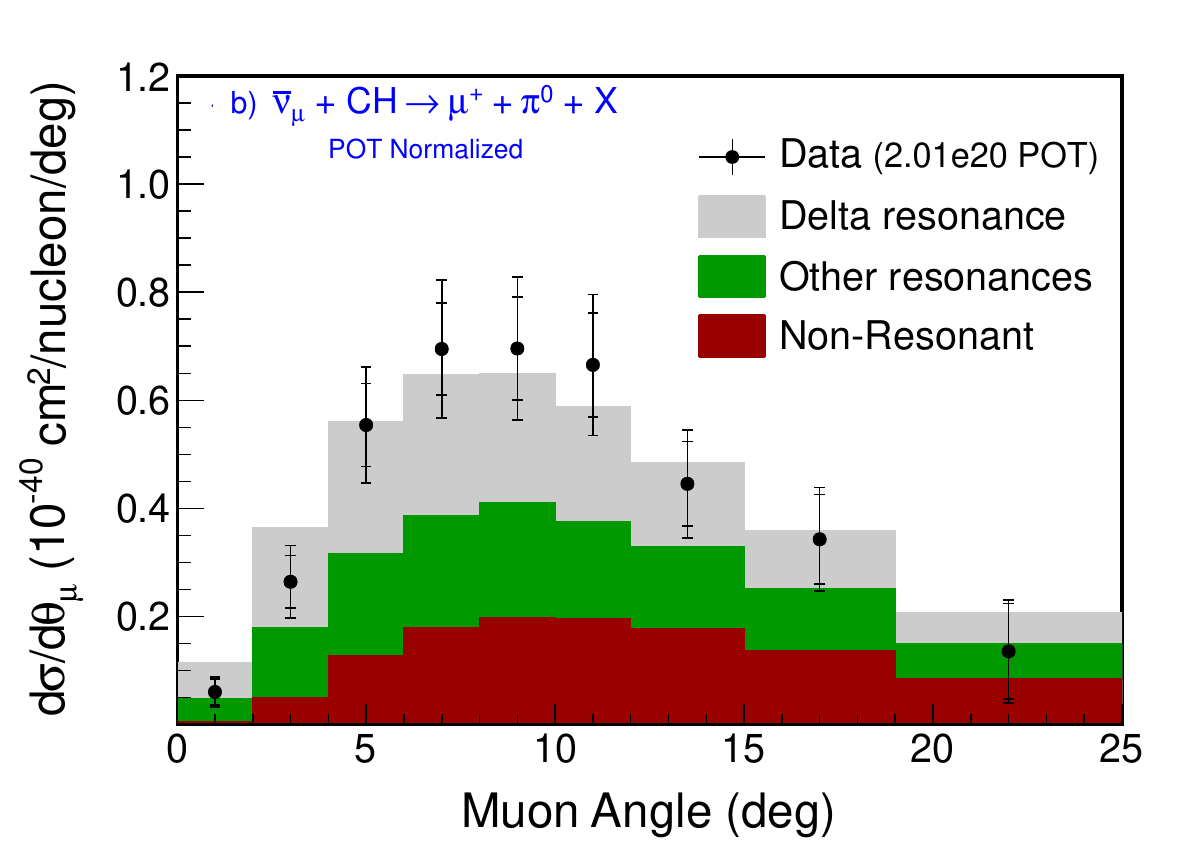}
                        }
{\caption{Differential cross sections $d\sigma/d\theta_{\mu}$ as in Fig. \ref{Fig06}
for the $\numu$ (a) and $\anumu$ (b) pion production samples, showing the
decomposition of the GENIE predictions into component reaction processes.   The three dominant
processes (histograms, bottom to top) are $\Delta(1232)$ production, higher-mass $N^{*}$ production, and 
pion non-baryon-resonance production.
\label{Fig08}}}
\end{center}
\end{figure}

\begin{figure}
  \begin{center}
              {
                \includegraphics[width=8.5cm]{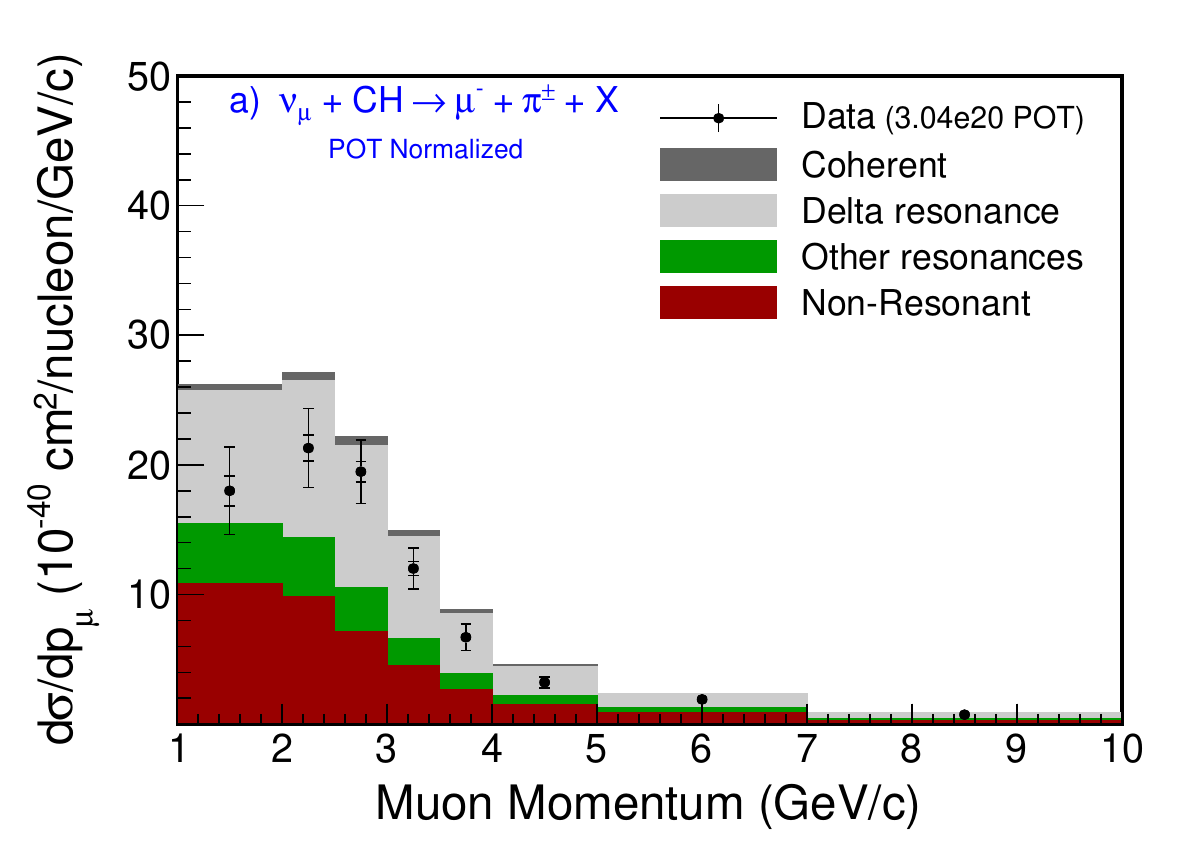}
              }
                        {
                          \includegraphics[width=8.5cm]{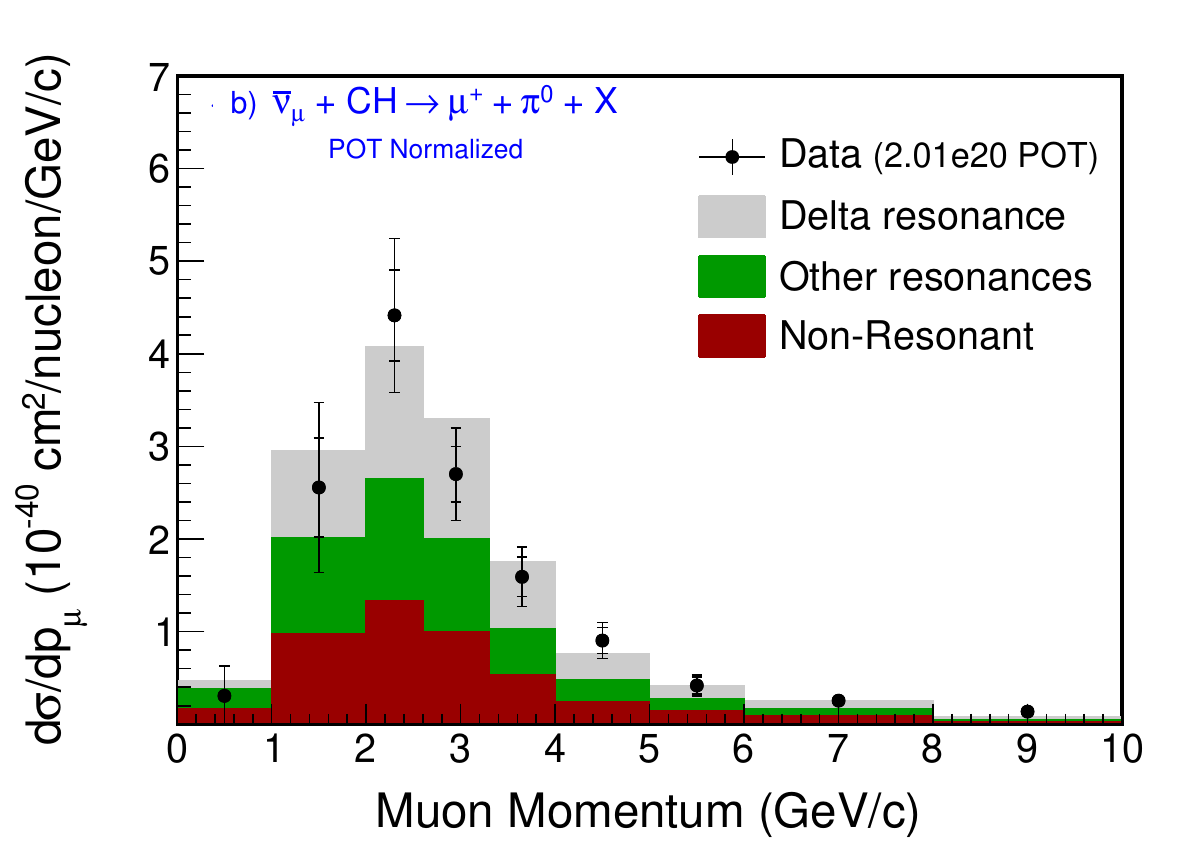}
                        }
{\caption{Differential cross sections $d\sigma/dp_{\mu}$ as in Fig. \ref{Fig07}
for the $\numu$ (a) and $\anumu$ (b) pion production samples, showing the
decomposition of the GENIE predictions into component reaction processes.   The component processes
are seen to distribute fairly uniformly with respect to muon momentum.
\label{Fig09}}}
\end{center}
\end{figure}

The event generators predict that several processes contribute to the event samples analyzed here.    The GENIE
prediction, for example, consists of quasielastic scattering, baryon resonance production, non-resonant pion production, 
DIS, and coherent pion production.    For the pion production samples
of this work, the topology selections and $W$ restriction ensure that the contributions from
quasielastic and DIS scattering are negligible.  The samples are predicted to be dominated by single-pion final states
arising from production and decay of the $\Delta(1232)$ and higher mass resonances, together with pion-nucleon non-resonant production.
Figures~\ref{Fig08} and \ref{Fig09} show 
the reaction-category composition for $d\sigma/d\theta_{\mu}$ and $d\sigma/dp_{\mu}$
of the $\numu$-CC($\pi^{+}$) and $\anumu$-CC($\pi^{0}$) samples.  
Referring to the component histograms, the $\numu$-CC($\pi^{+}$) sample 
(Figs.~\ref{Fig08}a and \ref{Fig09}a)
is estimated by GENIE to be comprised $\approx 50\%$ of $\Delta^{+,++}$ production, followed by 
non-resonant pion production, and production of higher mass $N^{*}$ states.
For the $\anumu$-CC($\pi^{0}$) sample (Figs.~\ref{Fig08}b and \ref{Fig09}b),
the $\Delta^{0}$ is also prominent, however 
the higher-mass $N^{*}$ contribution exceeds non-resonant pion production.   

A recent paper~\cite{Rodrigues-genie:2015} 
presents a new fit to the reanalyzed $\nu_\mu$-deuterium pion
production data~\cite{Wilkinson}.  
The best fit for GENIE 2.6.2 produces an increase in the resonant strength of
about 15\% and the non-resonant strength was decreased by about 50\%.
Those changes would produce better agreement between GENIE and the
charged pion data shown in Figs.~\ref{Fig08}a and \ref{Fig09}a.

The $\numu$-CC($\pi^{+}$) sample receives a small contribution from CC coherent single pion production.
The GENIE prediction for this contribution is shown by the component at small $\theta_{\mu}$ values in 
Fig.~\ref{Fig08}a (top, dark-shade histogram).  
Recall that in CC coherent pion production, the quantum (e.g. a pomeron) 
transferred to the struck nucleus carries no quantum numbers.    
The possible CC($\pi$) coherent reactions are:
\begin{equation}
\label{reaction-11}
 \numu (\anumu) + \mathcal{A} \rightarrow \mu^{-}(\mu^{+}) + \pi^{+}(\pi^{-}) + \mathcal{A}~.
\end{equation}

\vskip 2pt
\noindent
For coherent scattering to occur, the muon-pion system must have zero electric charge (like the incident $\nu/\anu$).
Thus coherent CC($\pi$) scattering is confined to the $\numu$-CC($\pi^{+}$) sample;
production of single $\pi^{0}$ mesons cannot occur via CC coherent  $\nu/\anu$ scattering.

Figure~\ref{Fig08}a shows coherent CC($\pi$) scattering to be
the only component process having a pronounced dependence on 
muon angle.   The three dominant processes are spread fairly uniformly 
over the angular range,  although the production of $\Delta(1232)$ is predicted to
gain prominence at very forward $\theta_{\mu}$ values.
On the other hand, Fig.~\ref{Fig09}a indicates that all of the component
processes, including coherent CC($\pi^{+}$) scattering, distribute broadly with respect to muon momentum.

\section{CC($\pi$) Cross Sections versus $E_{\nu}$}

\begin{figure}
  \begin{center}
              {
                \includegraphics[width=8.5cm]{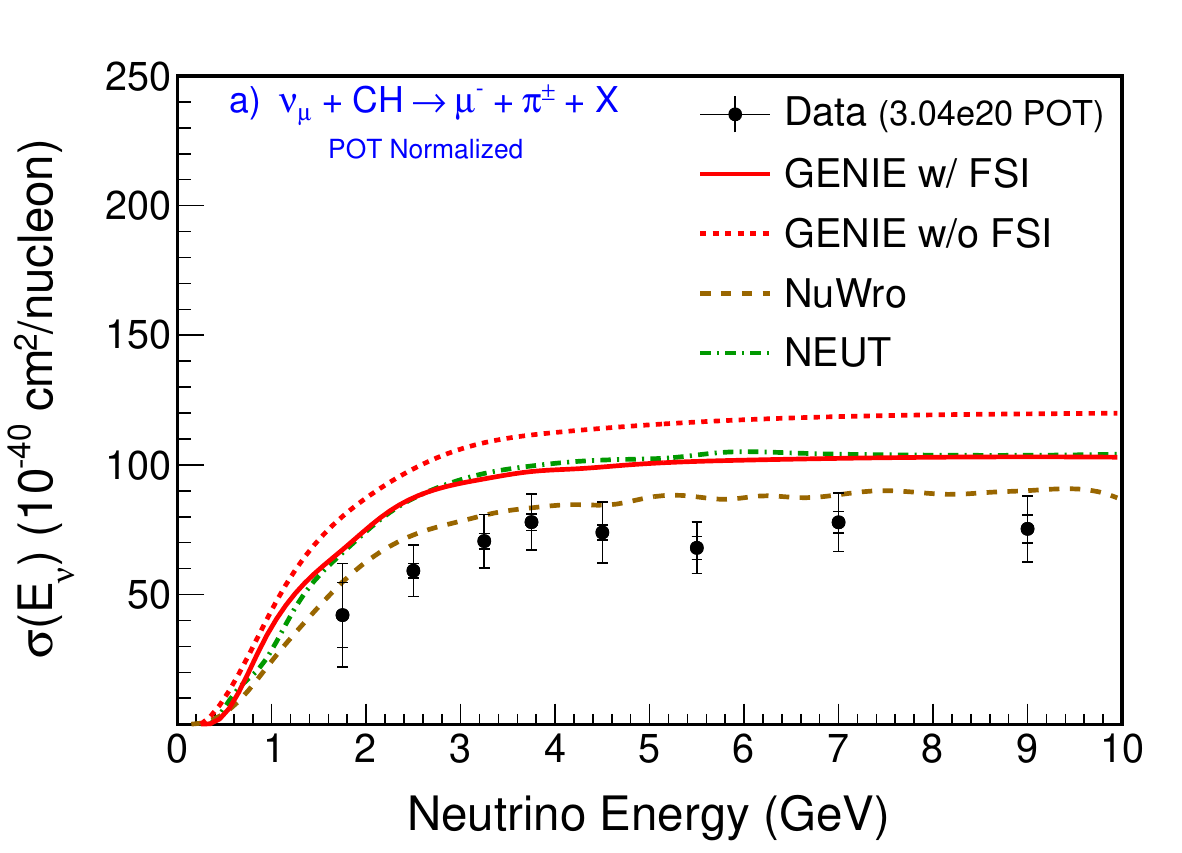}
              }
                        {
                          \includegraphics[width=8.5cm]{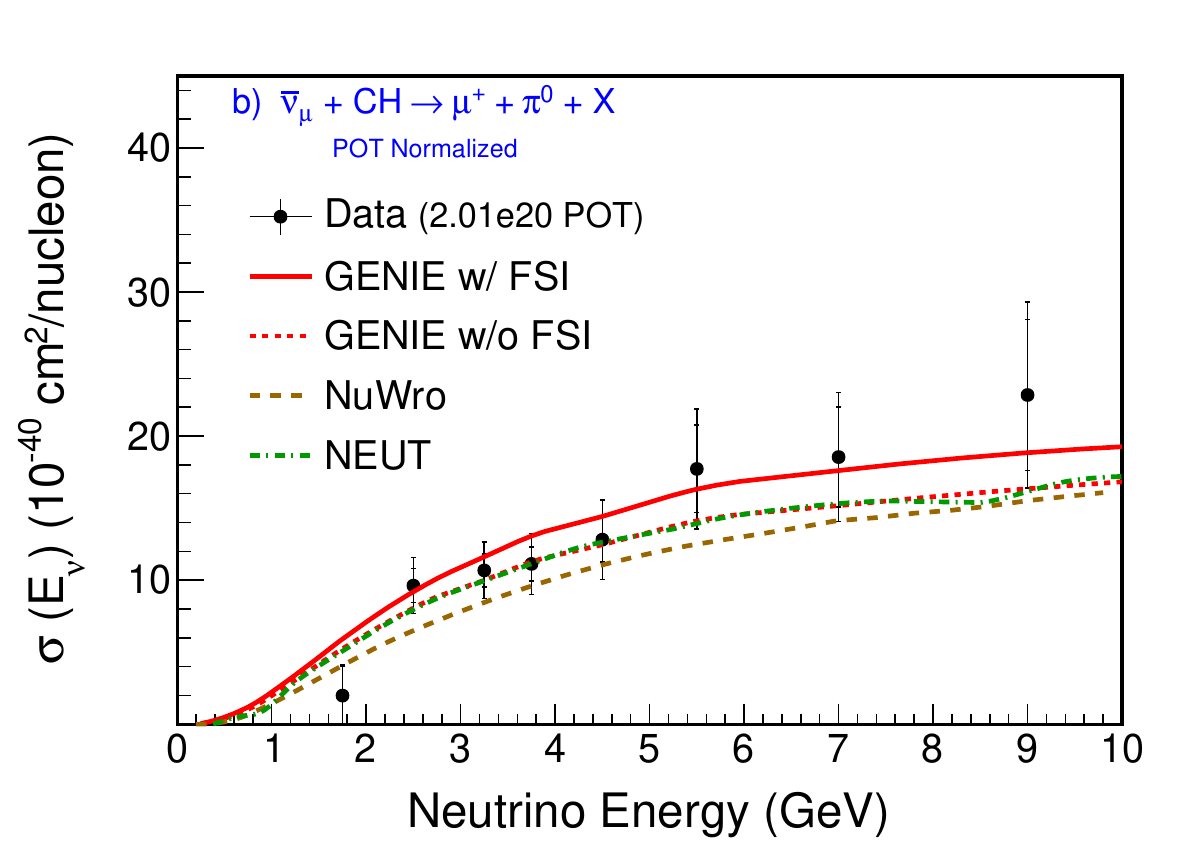}
                        }
\caption{Cross sections as a function of the neutrino energy 
         $E_\nu$  for the $\numu$ (a) and $\anumu$ (b) pion production samples.
	 Data are shown as solid circles.  The inner (outer) error bars correspond to statistical (total) uncertainties.
          The solid (dashed) distributions show GENIE predictions with (without) FSI, the long-dashed distribution is the
          prediction from the NuWro generator, and the dot-dashed distribution is the prediction from NEUT generator.}
\label{Fig10}
\end{center}
\end{figure}

Figure~\ref{Fig10} shows the cross sections as functions of neutrino (antineutrino) energy 
for the $\numu$-CC($\pi^{+}$) sample (upper) and for the $\anumu$-CC($\pi^{0}$) sample (lower plot).
Worthy of note is the difference in the ordinate ranges for the two plots.  For
the highest $E_{\nu}$ bin measured in each sample ($\langle E_{\nu} \rangle$ = 9.0 GeV), 
the cross section for  $\numu$-CC($\pi^{+}$) is more than twice as large as the $\anumu$-CC($\pi^{0}$) cross section. 
Also clearly discernible is the difference in the cross section rise-with-$E_{\nu}$ for the two samples.   
The cross section for $\numu$-CC($\pi^{+}$) sample (Fig.~\ref{Fig10}a)
reaches its plateau at $E_{\nu} \geq 3.0$ GeV.  However,  the cross section 
for $\anumu$-CC($\pi^{0}$) (Fig.~\ref{Fig10}b), 
exhibits a gradual rise throughout the measured region 1.5 $\leq E_{\nu} \leq 10.0$ GeV.   

The relative trends are a manifestation of the underlying vector minus axial-vector ($V-A$) structure of the hadronic currents
of these semileptonic weak interactions.    Within the structure functions of antineutrino CC scattering, the $V-A$ interference
terms have opposite sign compared to corresponding terms in the structure functions of neutrino CC scattering.   
The $V-A$ terms interfere destructively in the hadronic currents of $\anumu$-CC scattering, whereas the interference
is constructive in $\numu$-CC interactions.   The interferences contribute significantly to the cross sections in the sub-GeV
to few GeV range of $E_{\nu}$ and they account for the different trends in evolution with $E_{\nu}$ observed 
in Fig.~\ref{Fig10}a,b~\cite{Rein:1980wg}.

Figure~\ref{Fig10} compares the measured cross sections to the predictions of GENIE, NEUT, and NuWro.
The predictions for all of these generators exceed the measured $\numu$-CC($\pi^{+}$) cross section, with GENIE and NEUT
exhibiting a much larger disagreement (Fig.~\ref{Fig10}a).    For the $\anumu$-CC($\pi^{0}$) cross
section (Fig.~\ref{Fig10}b), there is less variation
among the generator predictions and much better agreement with the data.

\begin{figure}
  \begin{center}
              {
                \includegraphics[width=8.5cm]{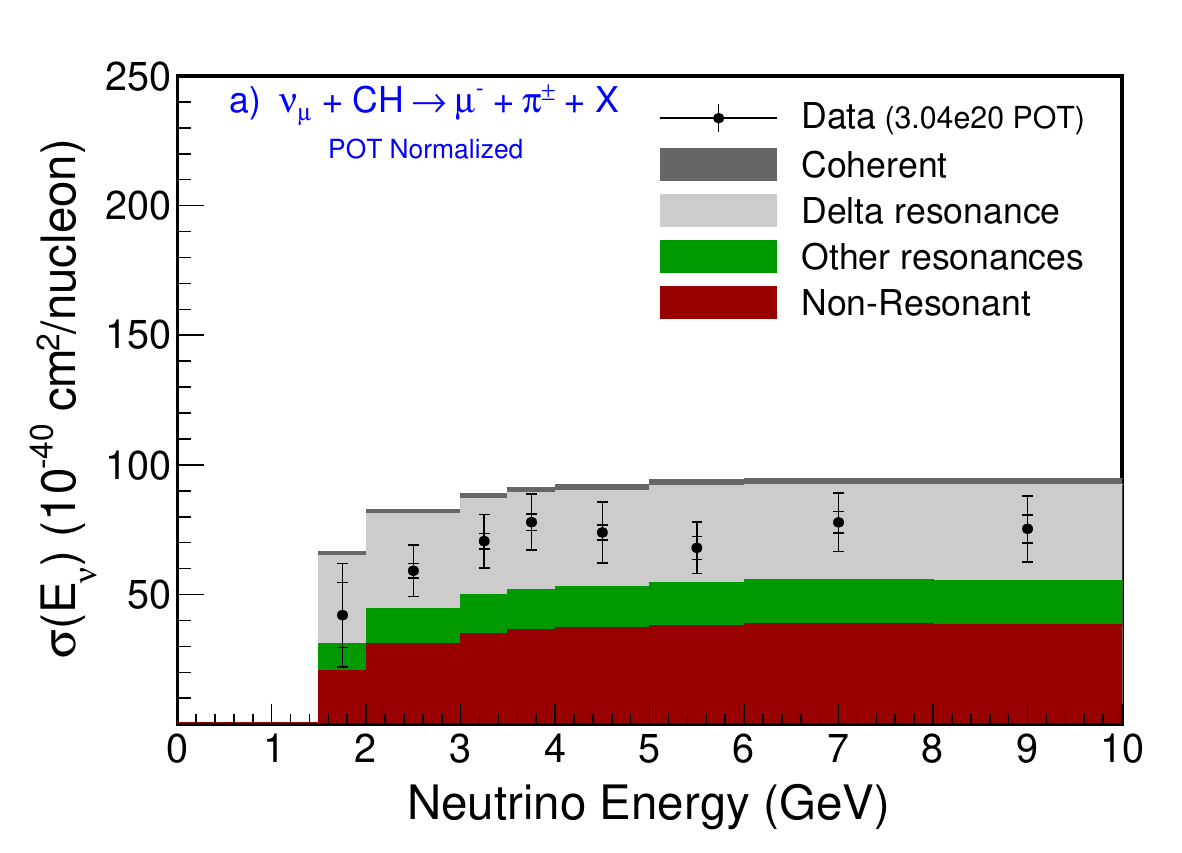}
              }
                        {
                          \includegraphics[width=8.5cm]{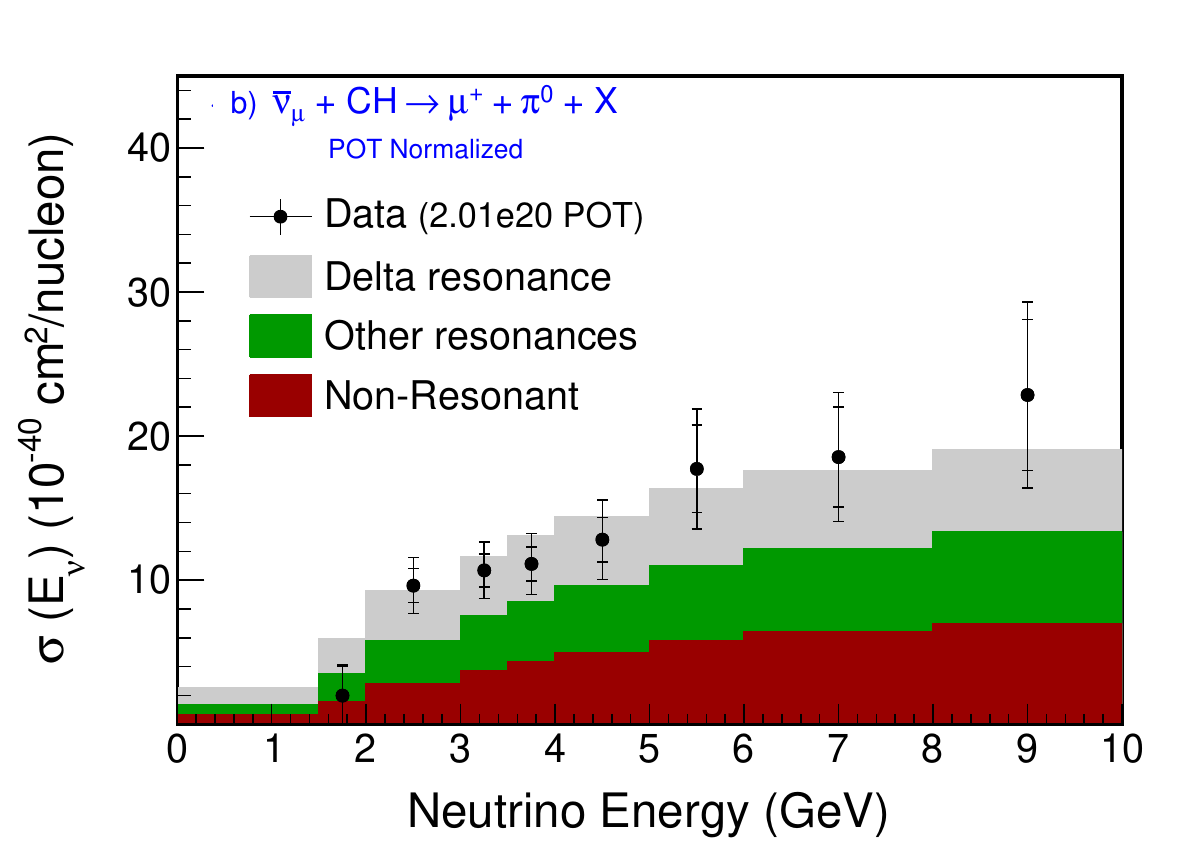}
                        }
{\caption{Component reaction processes in
       GENIE cross-section predictions for the $\numu$-CC($\pi^{+}$) (a)
       and $\anumu$-CC($\pi^{0}$) (b) samples.  Stacked histograms (bottom to top) 
       show the contributions from {\it (i)} pion 
       non-resonance processes, {\it (ii)} $N^{*}$ states above the $\Delta(1232)$,
       and {\it (iii)} $\Delta(1232)$ resonance production.
\label{Fig11}}}
\end{center}
\end{figure}

Figures~\ref{Fig11}a,b show the component reaction processes that are included in the GENIE
predictions for cross sections of the $\numu$-CC($\pi^{+}$) and $\anumu$-CC($\pi^{0}$) samples, respectively.
Notably absent are dramatic changes in the mixture of components 
with increasing $E_{\nu}$.  Although the $\Delta(1232)$ resonance 
is expected to dominate at low $E_\nu$ in all models, 
its relative contribution would be expected to decrease 
at higher $E_\nu$ where more energy is available to excite the struck nucleon.
The $W$ cut at 1.8 GeV however mitigates such an effect.
The pion non-resonant processes feature prominently in the GENIE predictions 
for both cross sections.  The separation into resonant and 
non-resonant processes is model dependent and could be different in other models.

\section{$d\sigma$/$dQ^2$ of CC($\pi$) Reactions}

\begin{figure}
  \begin{center}
              {
                \includegraphics[width=8.5cm]{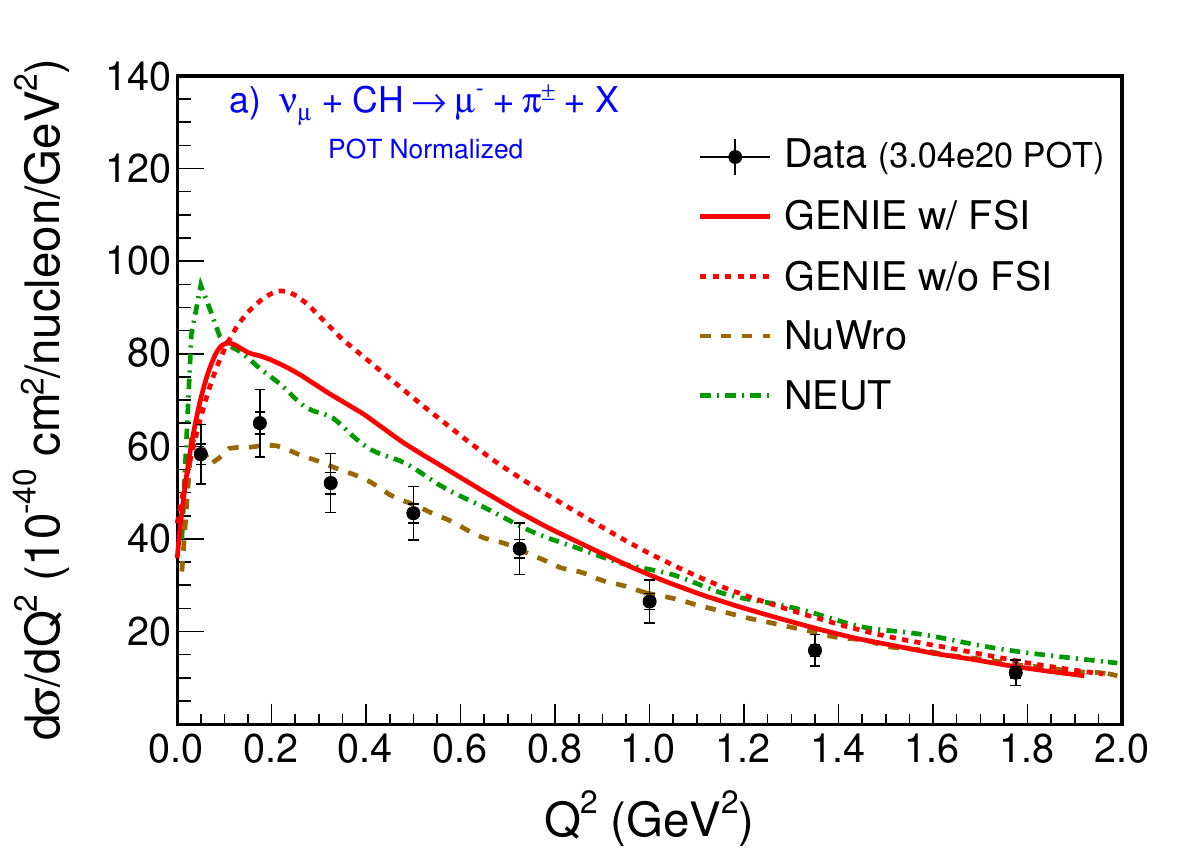}
              }
                        {
                          \includegraphics[width=8.5cm]{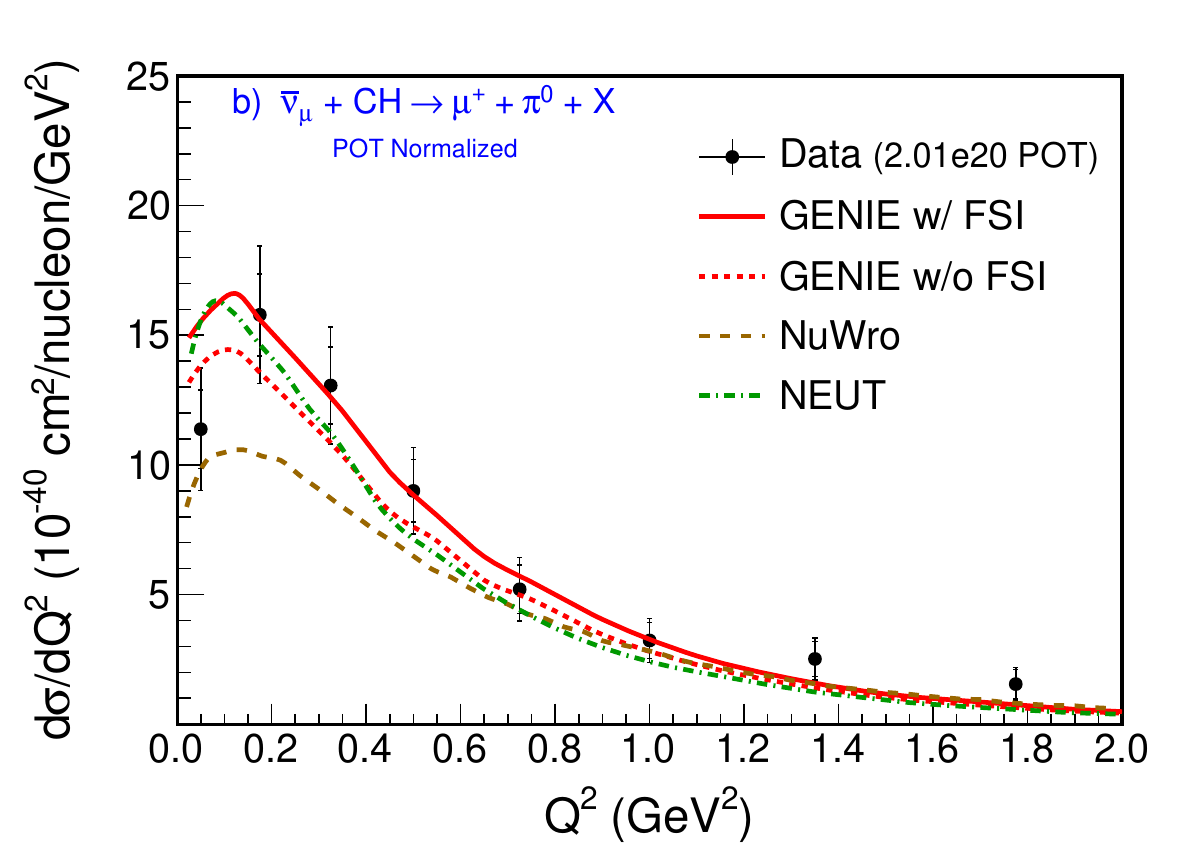}
                        }
\caption{Differential cross sections in four-momentum transfer squared $Q^2$ 
for the $\numu$-CC($\pi^{+}$) sample (a) and the $\anumu$-CC($\pi^{0}$) sample (b). 
Data are shown as solid circles.   The solid (dashed) distributions are GENIE predictions with (without) FSI, shown
together with predictions from the NuWro and NEUT event generators.
Ordinate-scale difference reflects the larger cross section for the $\numu$-CC($\pi^{+}$) sample.}                       
\label{Fig12}
\end{center}
\end{figure}

The differential cross sections as a function of $Q^2$ for the $\numu$-CC($\pi^{+}$) 
and $\anumu$-CC($\pi^{0}$) samples are shown in Fig.~\ref{Fig12}.
Note the large difference in the ordinate scales for the two distributions in corresponding $Q^2$ bins. 

For the generator predictions displayed in Fig.~\ref{Fig12}, NEUT and GENIE  
use a relativistic global Fermi gas model for nucleon momentum, while NuWro uses a local Fermi gas model.
The three calculations have very similar shapes for $Q^2>$ 0.2 GeV/c$^2$.  
At the lowest $Q^2$, it is possible that nucleon-nucleon correlations and Pauli blocking may contribute.  
These effects have been studied theoretically and experimentally in quasielastic neutrino scattering~\cite{martinimec,valenciamec}.

\begin{figure}
  \begin{center}
              {
                \includegraphics[width=8.5cm]{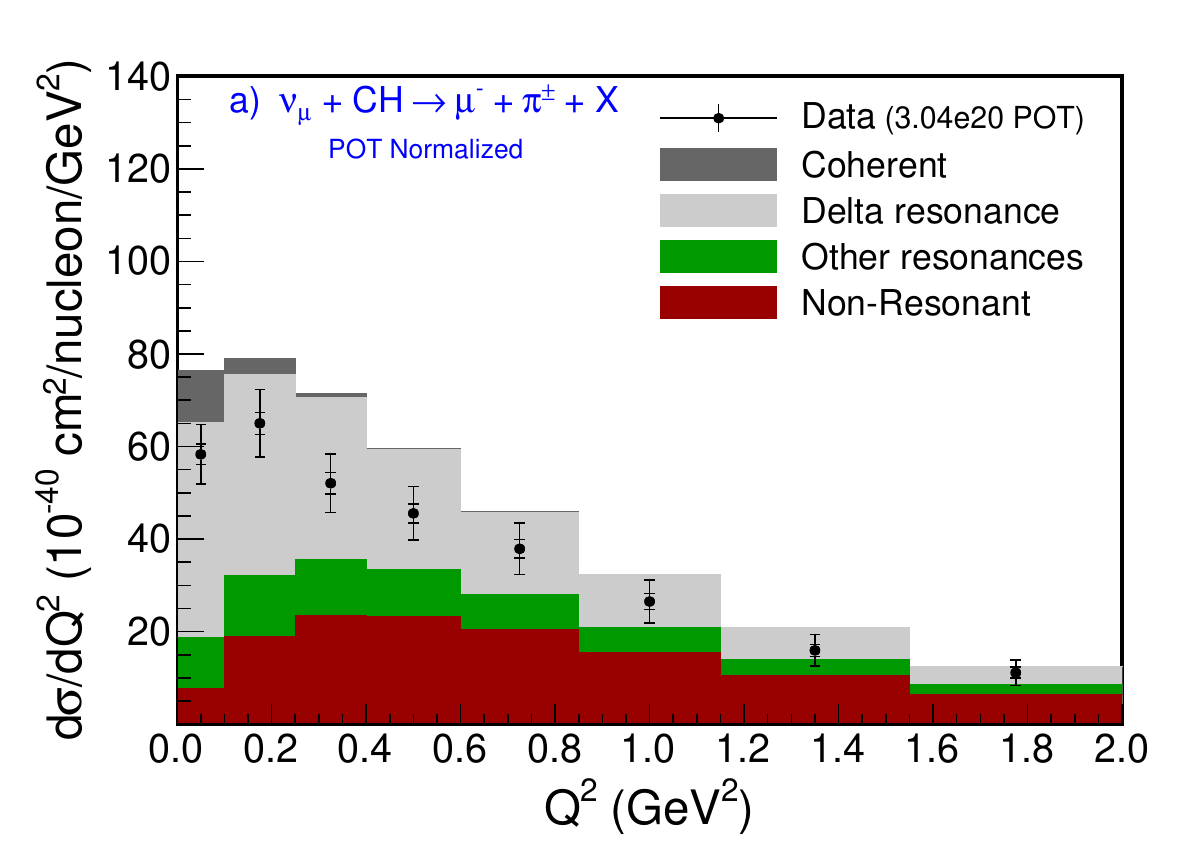}
              }
                        {
                          \includegraphics[width=8.5cm]{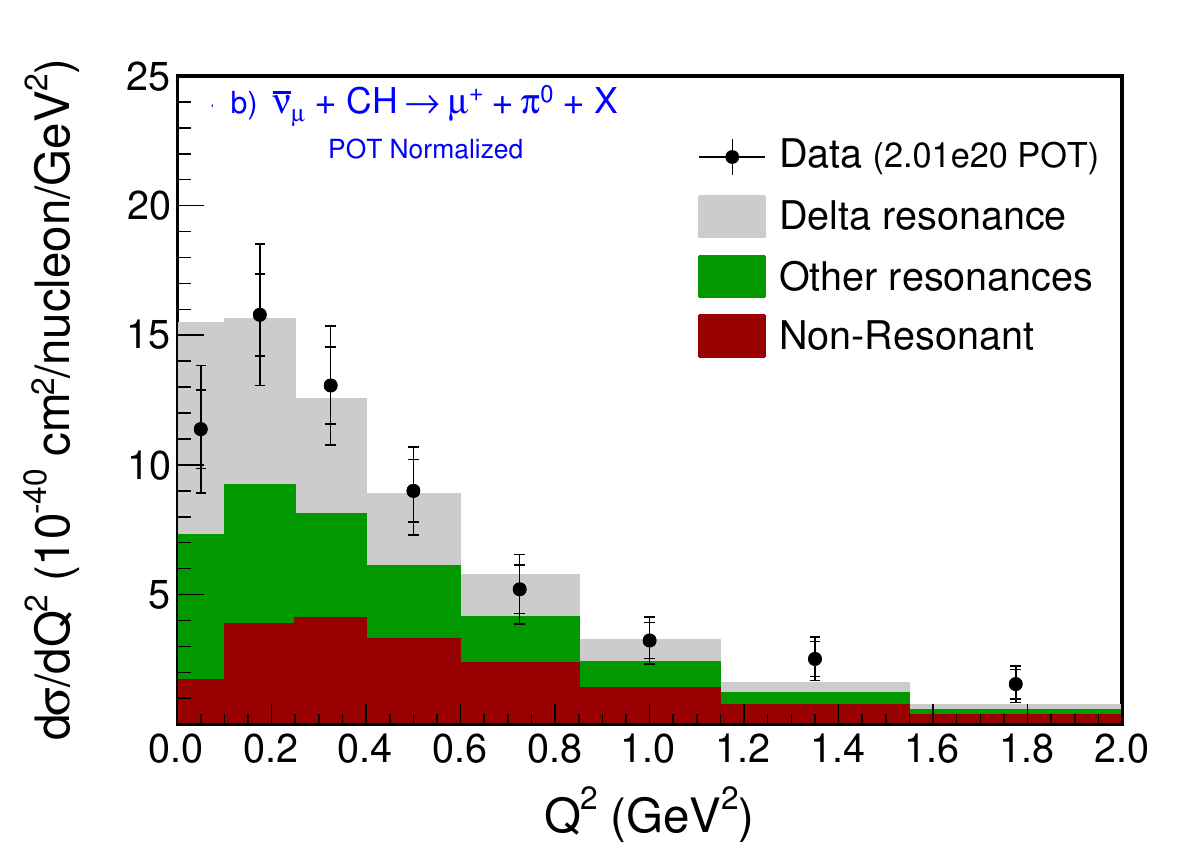}
                        }
{\caption{GENIE reaction processes for the $d\sigma/dQ^{2}$ data distributions 
for the $\numu$-CC($\pi^{+}$) (a) and $\anumu$-CC($\pi^{0}$) (b) samples. 
The coherent scattering contribution to $\numu$-CC($\pi^{+}$) is localized at very low $Q^{2}$.
\label{Fig13}}}
\end{center}
\end{figure}

Other effects can modify the cross section in $Q^{2}$ bins below 0.20 GeV$^2$.
Recall that coherent scattering can contribute to $\numu$-CC($\pi^{+}$) but not to $\anumu$-CC($\pi^{0}$).
By its nature, coherent scattering involves a very small four-momentum transfer to the target nucleus 
and so its contribution is confined to very low $Q^2$.  Different models
are commonly used; while NEUT and GENIE use different implementations
of the Rein-Sehgal~\cite{Rein:2006di} model, NuWro uses the Berger-Sehgal~\cite{Berger-coher:2009} model.
NEUT predicts a distinctly larger rate for coherent reaction \eqref{reaction-11} than does GENIE.
Consequently the NEUT prediction (Fig.~\ref{Fig12}a)
peaks near $Q^{2} \simeq 0.0$ GeV$^2$, while GENIE and NuWro do not predict such an effect.  In fact, GENIE and
NuWro predict a mild turnover in $d\sigma/dQ^{2}$ as $Q^{2}$ approaches zero GeV, in agreement with the
turnover exhibited by the data.  \minerva has published total cross section
data for coherent pion production~\cite{ref:CC-Coh-Minerva} using the same initial data samples as the analyses presented here.  
The NEUT prediction for the total coherent cross section is much larger than those data, 
while the GENIE prediction roughly agrees with the measured cross section in both shape and absolute rate.

Figure~\ref{Fig13} shows the GENIE predictions for the sample compositions compared to 
the $d\sigma/dQ^{2}$ data points.   The three main reaction categories are predicted to distribute broadly over the range
$0.0 \leq Q^{2} \leq 2.0$ GeV\,$^2$.   The coherent scattering contribution to the $\numu$-CC($\pi^{+}$) sample is predicted to be mostly confined
to $Q^{2} < 0.4\,$GeV$^2$.  At high $Q^2$, the non-resonant processes in GENIE have a
larger contribution to the data than the baryon resonance processes.   In Figs.~\ref{Fig13}a,b the GENIE predictions
exhibit the same trends as observed in Figs.~\ref{Fig08}, \ref{Fig09}, 
and \ref{Fig11}, namely good agreement with the $\anumu$-CC($\pi^{0}$) data   
and an absolute normalization that is high relative to the $\numu$-CC($\pi^{+}$) data.

\begin{figure}
  \begin{center}
              {
                \includegraphics[width=8.5cm]{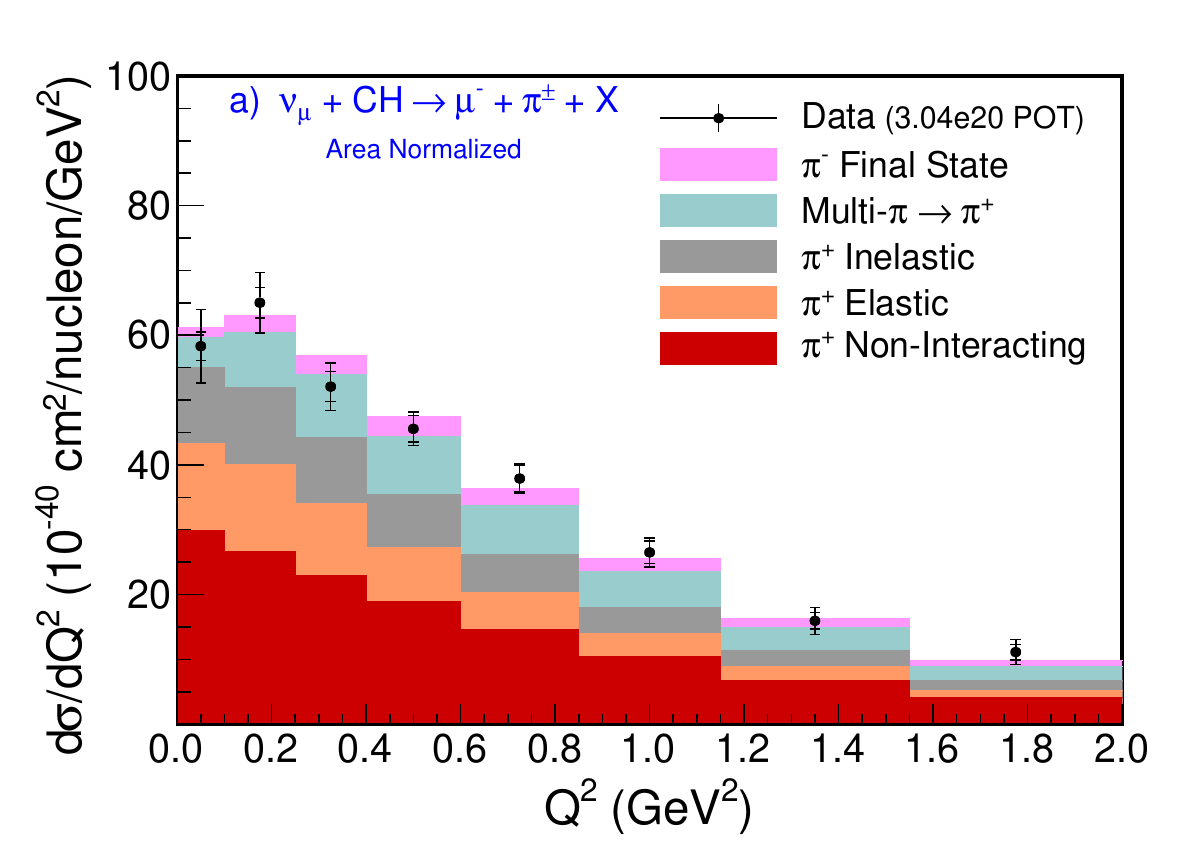}
              }
                        {
                          \includegraphics[width=8.5cm]{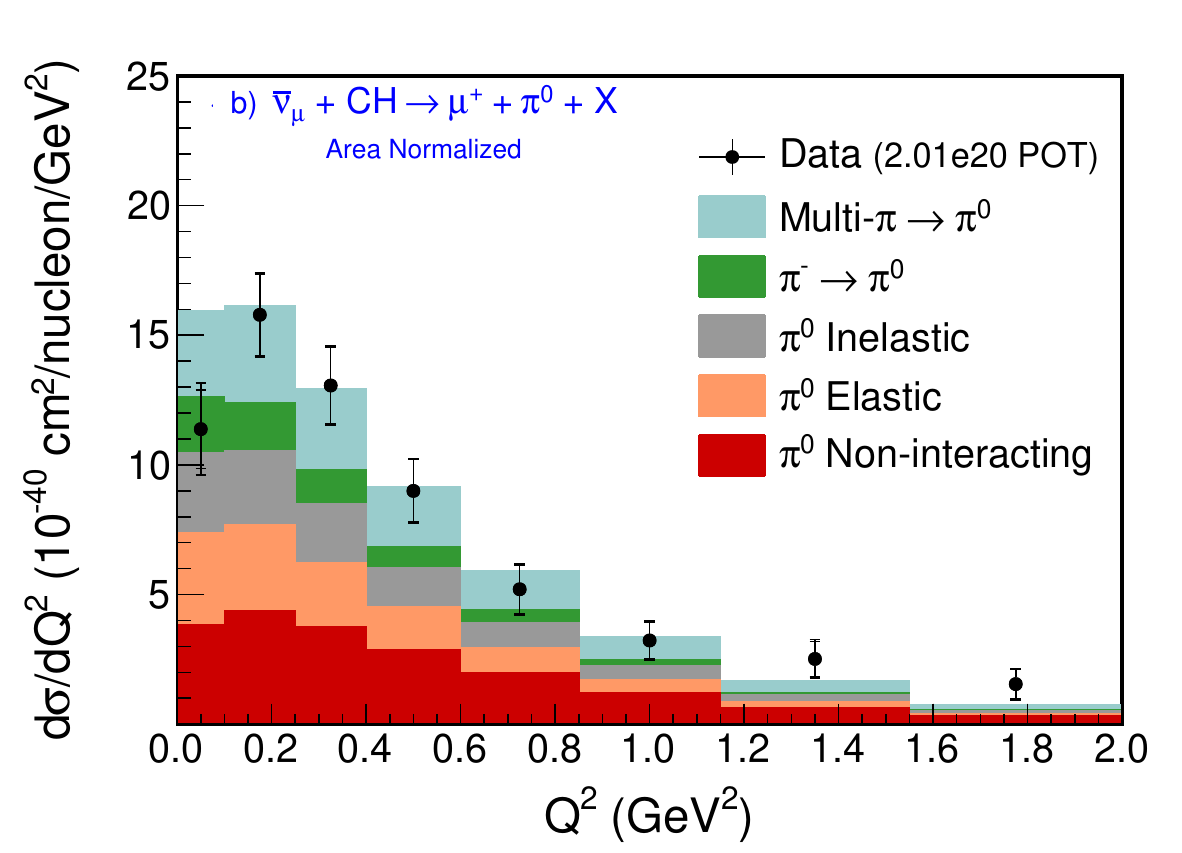}
                        }
{\caption{Breakout of pion FSI processes within
GENIE cross-section predictions for the two analysis samples.  
Stacked histograms (top to bottom) show contributions 
from $\pi^-$ (for the charged pion sample), multiple-pion production with absorption, 
charge exchange (for the $\piz$ sample), inelastic scattering, elastic scattering, and no interaction.
The full simulation, shown area-normalized to the data, reproduces the shape
of data $d\sigma/dQ^{2}$ without ascribing $Q^2$-dependence to pion FSI.
\label{Fig14}}}
\end{center}
\end{figure}

Figure~\ref{Fig14} shows a decomposition of the GENIE predictions 
according to pion FSI processes.    Here, the simulation has been area-normalized
to the data to better examine whether shape effects are present that could be
related to FSI.    The information displayed is 
complementary to that shown in Fig.~\ref{Fig13}.
While Fig.~\ref{Fig13} gives the GENIE subdivision by the primary process,
Fig.~\ref{Fig14} shows the subdivision by what happens after the primary process.
For the pion kinematic variables~\cite{Brandon-pion,Trung-pion}, the
FSI processes are important in determining the shape of the distributions.
However, for variables that are largely determined by
form-factor dependence and the nuclear model (such as $Q^2$), pion FSI processes
are expected to have a relatively mild affect.    Referring to Fig.~\ref{Fig12},
the GENIE-based predictions that omit or include FSI (dotted versus solid curves),
indicate that FSI processes reduce the differential cross section for $\numu$-CC($\pi^{+}$) and
elevate the differential cross section for $\anumu$-CC($\pi^{0}$).    The same trends are 
predicted by GENIE of the muon kinematic distributions 
of Figs.~\ref{Fig06} and \ref{Fig07},    
and for the cross sections as functions of neutrino energy in Fig.~\ref{Fig10}.

The data for $\anumu$-CC($\pi^{0}$) in Figs.~\ref{Fig12}b, 
\ref{Fig13}b, and \ref{Fig14}b 
(where coherent pion production cannot contribute) suggest a stronger turnover 
near $Q^{2} \simeq 0.0$ than is predicted by the generators.  This could be 
due to long-range nucleon-nucleon correlations,
usually treated via the Random Phase Approximation (RPA)~\cite{martinimec,valenciamec}, or to Pauli blocking.  
Pauli blocking should be applied for nonresonant pion production. 
Pion production through an intermediate $\Delta(1232)$ resonance should be 
subject to Pauli blocking, as $\Delta(1232)$ decay deposits a nucleon into the
residual nucleus.  The net effect is suppression of reactions at very low 
$Q^2$, similar to the suppression observed with neutrino quasielastic scattering.
The Pauli blocking effect has been calculated for $\mu^{\pm}\Delta(1232)$ channels produced in 
carbon-like nuclei in Ref.~\cite{Paschos-Pion-1}; the suppression is confined to $Q^2 < 0.2$ GeV$^2$ and to $W < 1.4$ GeV.   
The generator models shown here do not include any of these effects for pion production.

\section{Conclusions}
Differential cross sections in muon kinematic variables $\theta_{\mu}$ and $p_{\mu}$ are reported and compared for 
pion production processes $\numu$-CC($\pi^{+}$) and $\anumu$-CC($\pi^{0}$) obtained by exposing hydrocarbon to 
$\numu$ and $\numubar$ beams having similar flux profiles.  
Measurements of muon production angle and momentum are used to extract cross sections
as functions of $E_{\nu}$ and of $Q^{2}$ for the pion production processes. 
Events with total hadronic mass $W<1.8$ GeV 
are selected to emphasize the baryon resonance region.
Together with the previous publications based on the 
same data sample~\cite{Trung-pion,Brandon-pion}, these measurements provide a 
broad picture of pion production for neutrino energies 1.5\,--\,10\,GeV 
wherein charged-current scattering from single nucleons is convolved with nuclear
structure and pion FSI effects.   Data summary Tables for the measurements of this work 
that may facilitate phenomenological investigations,
are available in the Supplement~\cite{Supplement}.    

The ensemble of differential cross sections are compared to simulations 
based upon the GENIE, NEUT, and NuWro event generators.    
For differential cross sections measured with the 
$\numu$-CC($\pi^{+}$) sample, the absolute event rates predicted by 
GENIE and NEUT are observed to exceed the measurements of this work by amounts that are
typically between 1-to-2\,$\sigma$ of the data. 
 GENIE and NEUT predictions give much better agreement 
for the $\anumu$-CC($\pi^{0}$) comparisons.  On the other hand, the NuWro generator obtains agreement with the 
distributions for the $\numu$-CC($\pi^{+}$) sample, but predicts event rates that are generally low by up to 1.5\,$\sigma$ relative to the data for 
$\anumu$-CC($\pi^{0}$).    The differences in generator predictions arise from differences 
in the cross section for pion production on free nucleons~\cite{Rodrigues-genie:2015}, 
and from the treatment of FSI, which has significant uncertainties.
In contrast to the assorted discrepancies for absolute rates,  
all three generators obtain respectable shape agreement with the full suite of differential cross sections.

The suite of differential cross sections for the two samples is examined 
in light of the reaction category composition used by the GENIE generator.
The likely role of coherent CC($\pi^{+}$) scattering 
in the $\numu$-CC($\pi^{+}$) sample is thereby illustrated.     
Comparisons of the three generators
with $d\sigma/dQ^{2}$ for this sample indicates that the rate 
of coherent CC($\pi^{+}$) is set too high in NEUT.  
The $d\sigma/dQ^{2}$ distribution of $\numu$-CC($\pi^{+}$) has sensitivity to nuclear structure.  
Neither Pauli blocking nor nucleon-nucleon 
correlations are included in the default options of the GENIE, NuWro, or NEUT versions used here.  Despite 
the simplicity of the nuclear models employed by these generators, 
the shapes of the $Q^2$ distributions predicted by GENIE and NuWro agree with the data.

In summary, the measurements and event sample comparisons of this work 
shed light on CC pion production
by neutrinos and antineutrinos.  A correct description of these
data requires accurate models for this multifaceted problem.  Separate 
understanding of the three aspects (pion production
from the bound nucleon, nuclear structure, and pion FSI) are important for interpreting
events recorded by the neutrino oscillation experiments at long baselines.
Fortunately, the complete data set allows for some 
separation of processes.   The measurements 
illuminate shortfalls in current generators with respect to absolute rate
predictions for coherent, resonant, and non-resonant CC production of pions.  These 
results can guide the development of improved 
neutrino-interaction models that are
important to the international effort to obtain precision measurements 
of neutrino oscillations using $\numu$ and $\numubar$ beams.

\section*{Acknowledgments}
 This work was supported by the Fermi National Accelerator Laboratory under US Department of Energy contract
No. DE-AC02-07CH11359 which included the MINERvA construction project.  Construction support was also granted by
the United States National Science Foundation under Award PHY-0619727 and by the University of Rochester.  
Support for participating scientists was provided by NSF and DOE (USA), by CAPES and CNPq (Brazil), by CoNaCyT (Mexico), 
by CONICYT (Chile), by CONCYTEC, DGI-PUCP and IDI/IGI-UNI (Peru), and by Latin American Center for Physics (CLAF).  
We thank the MINOS Collaboration for use of its near detector data.  We acknowledge the dedicated work of the Fermilab 
staff responsible for the operation and maintenance of the beamline and detector, 
and we thank the Fermilab Computing Division  for support of data processing.

\section*{Appendix: Pion Kinematic Distributions}

\begin{figure}
  \begin{center}
              {
                \includegraphics[width=8.5cm]{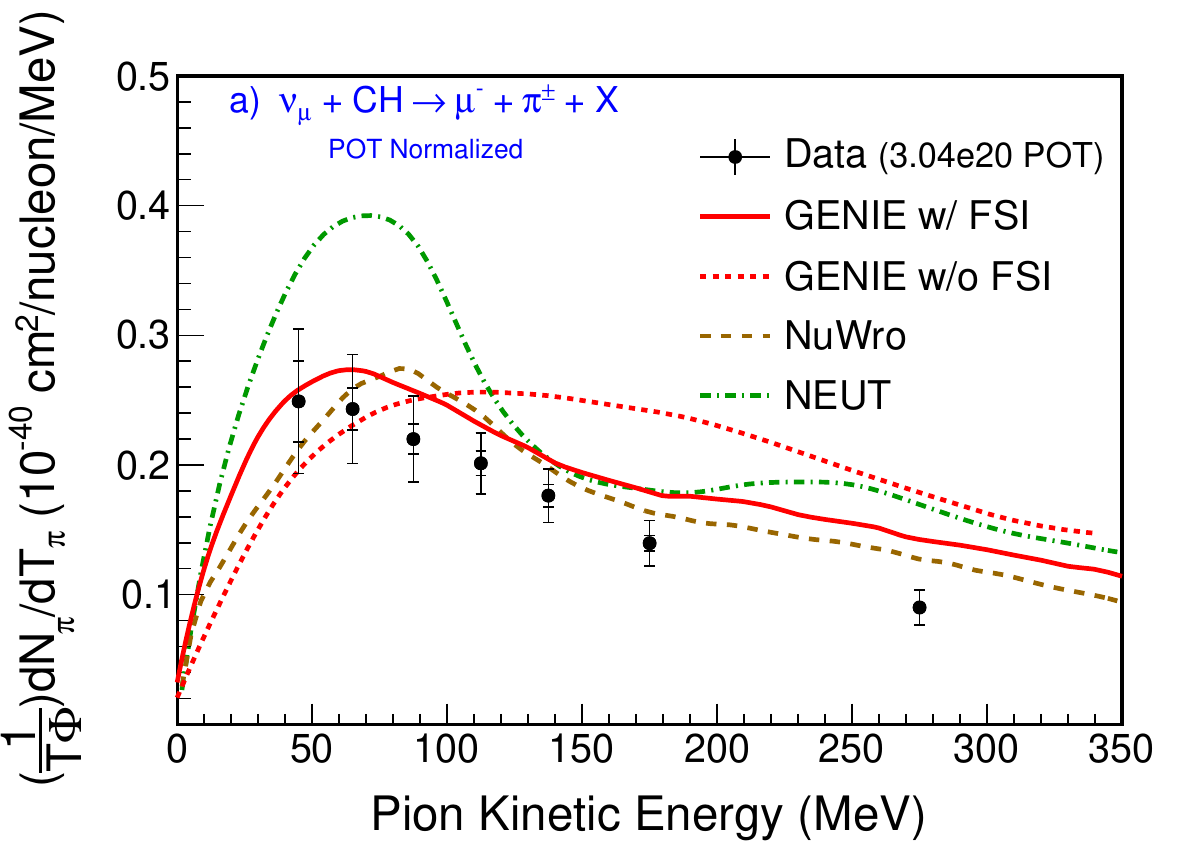}
              }
                        {
                          \includegraphics[width=8.5cm]{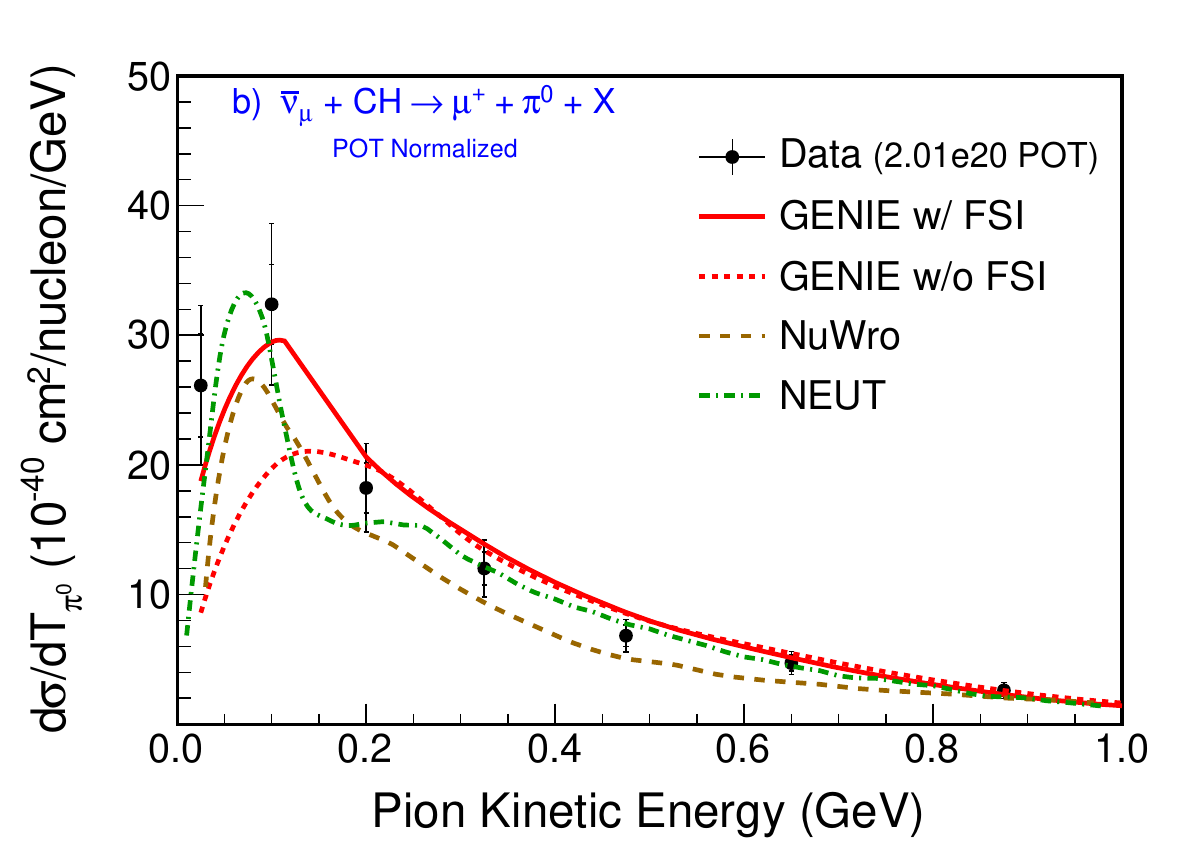}
                        }
{\caption{Differential cross sections for pion kinetic energy, $d\sigma/dT_{\pi}$, for the $\numu$-CC($\pi^{+}$) (a)
and $\anumu$-CC($\pi^{0}$) (b) samples.   The data (solid circles) are compared to GENIE predictions
neglecting versus including pion FSI (dashed vs solid-line histograms).   Improved descriptions for  
shapes of the pion spectra are obtained with FSI effects included in the simulations. 
\label{Fig15}}}
\end{center}
\end{figure}

Distributions for $\mu^{\pm}$ and related 
kinematic variables are featured by the main text.
This Appendix presents distributions describing pion production kinematics for the two
analysis samples.  These figures represent updates to similar plots presented in the previous papers~\cite{Brandon-pion,
Trung-pion}, reflecting the improvements in the neutrino and antineutrino flux estimates noted in Sec.~\ref{subsec:B-D-E}.

\begin{figure}
  \begin{center}
              {
                \includegraphics[width=8.5cm]{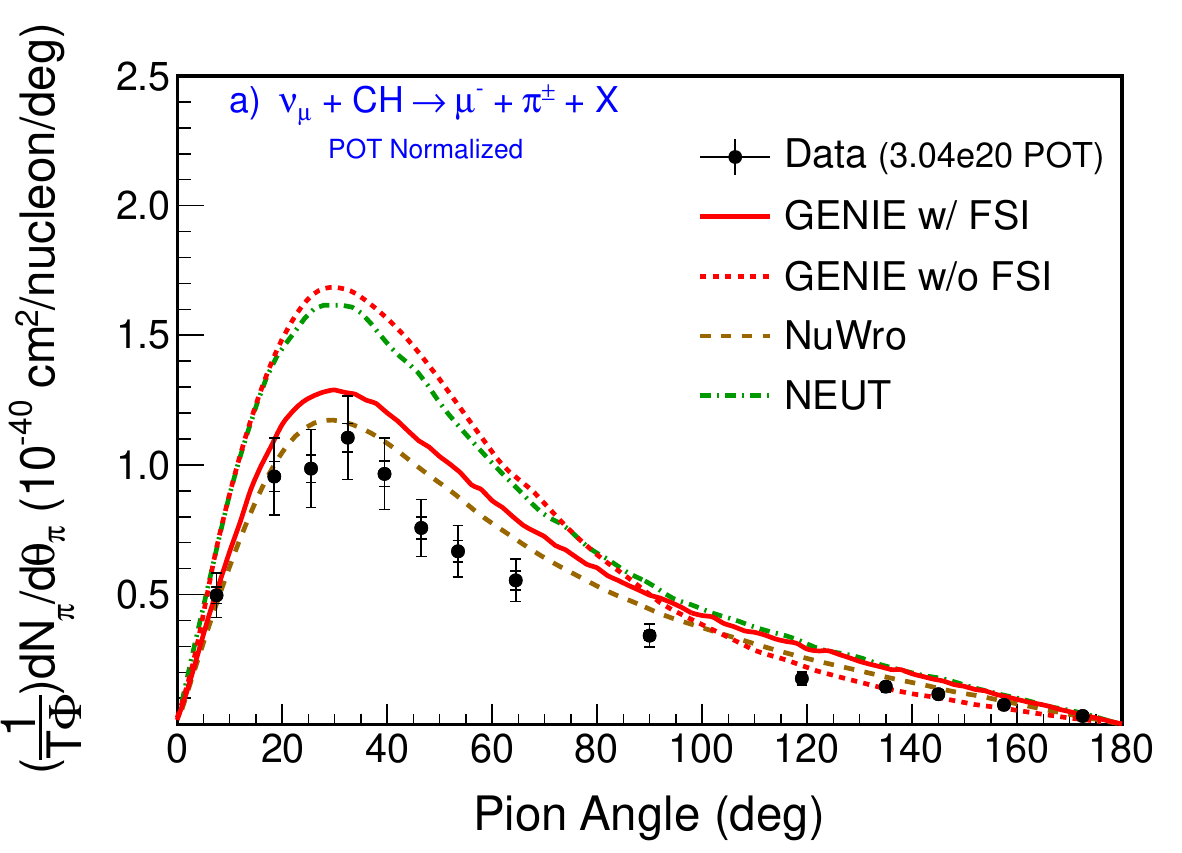}
              }
                         {
                          \includegraphics[width=8.5cm]{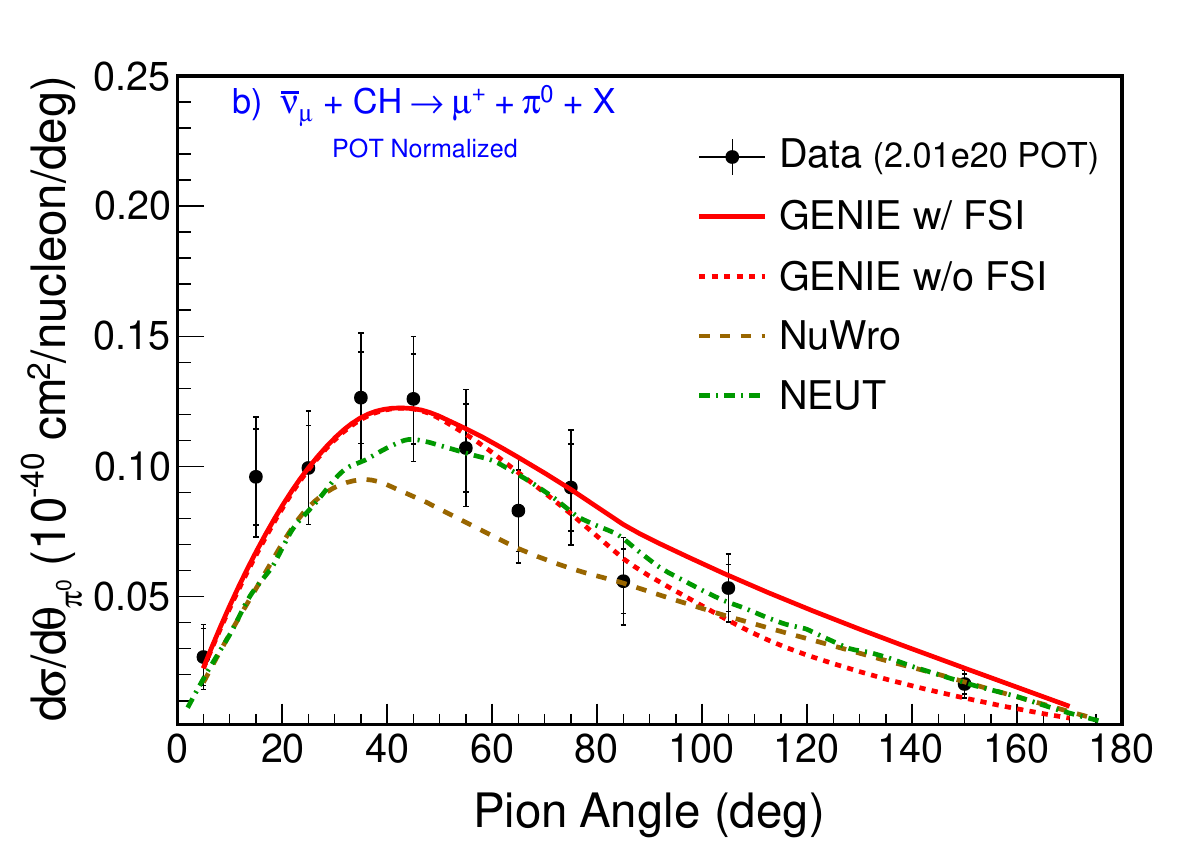}
                        }
{\caption{Differential cross sections for the pion production angle with respect to the beam direction,
$d\sigma/d\theta_{\pi}$, for the $\numu$-CC($\pi^{+}$) (a) and $\anumu$-CC($\pi^{0}$) (b) samples.  
As in Fig. \ref{Fig15}, the data is compared to GENIE predictions without and with pion FSI
effects included;  marked improvement with the data is observed when pion FSI is taken into account (solid-line distributions).
\label{Fig16}}}
\end{center}
\end{figure}

Figure~\ref{Fig15} shows the flux-averaged pion kinetic energy for
the charged pions of the $\numu$-CC($\pi^{+}$) sample (Fig.~\ref{Fig15}a), and for the $\piz$ of the 
$\anumu$-CC($\pi^{0}$) sample (Fig.~\ref{Fig15}b).  
The kinetic energy ranges of $\pi^+$ and $\pi^0$ 
are different because the maximum $\pi^+$ energy is closely related to the detector depth.  
Figure~\ref{Fig16} shows the polar-angle distributions for the produced $\pi^{\pm}$ and
for the $\piz$ of these samples.
The data points depict the same signal obtained with the same procedures reported in the main text,
including the restriction on the invariant hadronic mass, $W < 1.8$ GeV.
For the $\anumu$-CC($\pi^{0}$) sample, the updated $\pi^0$ distributions are shown 
for the same energy range, 1.5\,GeV $< E_\nu < 10$\,GeV, as is used 
for the $\numu$-CC($\pi^{+}$) sample. The quantity plotted for charged pions
in Figs.~\ref{Fig15}a and \ref{Fig16}a is same 
as in Ref.~\cite{Brandon-pion}.  Events with one or more $\pi^{+}$ or $\pi^{-}$ are included.
Although not a true cross section,
it arises from the cross section definition, Eq.~\eqref{eq:dif-xsec}, 
when each event can produce one or more pions.  Each charged pion
contributes one entry to a histogram and the cross section is calculated as in Sec.~\ref{X-sec-calc}.

Figures~\ref{Fig15} and~\ref{Fig16} show comparisons with GENIE, NEUT, and NuWro predictions;
for both samples it is clearly seen that, for GENIE, the pion FSI treatment causes the simulation to move closer to the data.
The main change in these updated results compared to the earlier ones is reduction of the large disagreement
in absolute normalizations of the differential cross sections between data and GENIE-based predictions.
This reduction comes about because the calculated data normalizations are now higher
by 13\% and 12\% for the $\numu$ and $\anumu$ event samples respectively as the 
result of revisions made to the flux estimates of the exposures as described in Sec.~\ref{X-sec-calc}.   (Note that all pion distributions shown here are obtained using the $W < 1.8$ GeV selection; this cut was not applied in the $\piz$ distributions of Ref.~\cite{Trung-pion}.) Figures~\ref{Fig15} and \ref{Fig16} show that the GENIE predictions are now closer to the data but still appear to be high, 
with the overall normalization difference exceeding $1\sigma$ in the $\numu$-CC($\pi^{+}$)  data set.


\end{document}